\documentclass[aip,reprint]{revtex4-1}
\draft 

\usepackage[english]{babel}
\usepackage[utf8]{inputenc}
\usepackage[T1]{fontenc}
\usepackage{amsmath}
\usepackage{amssymb}
\usepackage{mathrsfs}
\usepackage{xspace}
\usepackage{bm}
\usepackage{array}
\usepackage{graphicx}
\usepackage{float}
\usepackage{subcaption}
\usepackage{url}
\usepackage{diagbox}
\usepackage{mydef,boldfonts}
\usepackage{color}

\newcommand{\reffig}[1]{Fig.~\ref{#1}} 
\newcommand{\refeq}[1]{Eq.~\eqref{#1}} 
\newcommand{\refeqs}[2]{Eqs.~\eqref{#1}~and~\eqref{#2}} 

\begin{document}


\title{Two- and Three-Dimensional Simulations of Rayleigh-Taylor Instabilities Using a Coupled Cahn-Hilliard / Navier-Stokes Model} 



\author{R. Zanella} 
\email{zanella.raphael@gmail.com}
\affiliation{Laboratoire PMC, Ecole Polytechnique, CNRS, IP Paris, 91128 Palaiseau Cedex, France}

\author{G. Tegze}
\affiliation{Wigner Research Centre for Physics, P.O. Box 49, H-1525 Budapest, Hungary}

\author{R. Le Tellier}
\affiliation{CEA, DES, IRESNE, DTN, Cadarache, F-13108 Saint Paul-lez-Durance, France}

\author{H. Henry}
\affiliation{Laboratoire PMC, Ecole Polytechnique, CNRS, IP Paris, 91128 Palaiseau Cedex, France}



\date{\today}

\begin{abstract}
We report on two- and three-dimensional numerical simulations of Rayleigh-Taylor instabilities in immiscible fluids. A diffuse-interface model that combines the Cahn-Hilliard equation, governing the evolution of the volume fraction of 
one fluid, and the Navier-Stokes equations, governing the bulk velocity and pressure, is used. 
The study is limited to low Atwood numbers owing to the use of the Boussinesq approximation. 
The code is based on a pseudo-spectral method. 
A linear analysis is first performed in a two-dimensional case of Rayleigh-Taylor instability to confirm that the model 
very well captures this phenomenon in the case of inviscid or viscid fluids. One key aspect of this work is that 
the influence of the thermodynamic parameters related to the Cahn-Hilliard equation (interface thickness and mobility) 
is quantitively studied. Three-dimensional results of Rayleigh-Taylor instabilities in viscous fluids are then presented 
to show the possibilities of this modeling. We observe the effect of the viscosity and the wavelength of an initial 
single-mode perturbation on the mass transport during the nonlinear regime.
\end{abstract}


\maketitle 


\section{Introduction}

The Rayleigh-Taylor instability is a hydrodynamic instability that occurs when two fluids are superimposed with 
the heavier one on top. If the difference of mass density (simply called density in the following) is sufficient to overcome 
surface tension, an infinitely small perturbation can 
destabilize the interface and lead to the progressive interpenetration of the fluid layers, until the heavier fluid has 
totally dived under the lighter one\cite{sharp_physica12D_1984}. This instability can be found in many natural 
phenomena (cosmology, geology, oceanography) or industrial applications (combustion, heat exchangers, mixing, 
aerosol transport)\cite{zhou_physicsReports_2017}. 
The problem has been deeply studied in the case of a single-mode perturbation in order to understand the various stages of the instability. As long as the amplitude of the perturbation is small compared to its wavelength, the growth of the perturbation is exponential\cite{taylor_rsl_1950}. The growth rate of the instability depends on the excited mode and the densities of the fluids, but also on the surface tension and the viscosity\cite{chandrasekhar_clarendon_1961,menikoff_pof_1977}. This regime, said linear because the nonlinear effects are negligible, is followed by a quasi-steady regime. Experiments\cite{lewis_rsl_1950, waddell_pof_2001, wilkinson_pof_2007} revealed that the velocity of the "bubble" of light fluid rising through the heavy fluid reaches a constant value after the exponential growth. Models of terminal velocity taking into account the excited mode and the fluid properties have been theoretically established and confronted to numerical simulations in 2D and 3D\cite{oron_pop_2001, goncharov_prl_2002, young_jt_2006, sohn_prE_2009}. After the two first regimes, a re-acceleration regime\cite{ramaprabhu_prE_2006, wei_prE_2012, ramaprabhu_pof_2012, hu_pof_2019} and a chaotic regime can develop\cite{ramaprabhu_pof_2012}.
However, most of the aforementioned work is focused on the high Reynolds number regime where inertia is dominant and little is known on the low or intermediate Reynolds number regime.

For this purpose, we have used a diffuse-interface model and more precisely a coupled Cahn-Hilliard / Navier-Stokes model (CHNS model).
Diffuse-interface models rely on the use of a transport equation for an order parameter
that indicates in which phase each point of space is and that smoothly varies between the 
phases\cite{anderson_arfm_1998}. As a result, the
interface is tracked implicitly as an iso-surface of the order parameter. The
boundary conditions at the interface are then imposed through a proper
coupling between the transport equation of the order parameter and the flow
equations. 
In the case of the CHNS model, the transport equation is the Cahn-Hilliard
equation\cite{cahn_jchp_1958, cahn_jchp_1965} and it can be coupled in a thermodynamically
consistent way with the Navier-Stokes equations so that capillary effects are well
described. 
However, non-hydrodynamic parameters related to the Cahn-Hilliard equation appear in the modeling: the interface 
thickness at thermodynamic equilibrium and the mobility. The use of a Cahn-Hilliard model in
practical situations involves the upscaling of the interface thickness from
actual sizes to sizes compatible with numerical
simulations (a small but not too small fraction of the characteristic lengthscale of
the system). The effects of this upscaling and how to perform it by appropriately choosing the interface thickness and the mobility is a delicate problem.
A recipe has recently been proposed in Ref.~\onlinecite{magaletti_jfm_2013}, suggesting a scaling between the mobility and the ratio of the interface thickness and the characteristic length of the problem.

The CHNS model has already been used to simulate Rayleigh-Taylor instabilities in various studies. In Ref.~\onlinecite{celani_jfm_2009}, the model was used in two dimensions and shown to give fairly accurate results in the linear and weakly nonlinear regimes for viscous and non viscous fluids. However, the convergence of the model with respect to the thermodynamic parameters (interface thickness and mobility) was not presented. In Refs.~\onlinecite{lee_ijnme_2011, lee_cma_2013}, two- and three-dimensional simulations were performed for physical parameters corresponding to moderate Atwood number and relatively high Reynolds number (but no surface tension). The simulations showed the ability of the model to qualitatively reproduce the expected behavior but the influence of the Péclet number (related to the mobility) was only qualitatively discussed in Ref.~\onlinecite{lee_ijnme_2011}. Ref.~\onlinecite{hong2019} focused on the chaotic regime at high Reynolds number. However, no discussion of the role of the thermodynamic parameters was provided. In Ref.~\onlinecite{nitschke_jfm_2012}, the Rayleigh-Taylor instability in a torus was used as a qualitative test of a finite element approach. No physical explanation was given regarding the thermodynamic parameters. Ref.~\onlinecite{yang2016} addressed the particular case of Rayleigh-Taylor instabilities under the influence of an electric field. The influence of the thermodynamic parameters was nevertheless not studied. The interplay of flow and diffusion during the mixing of miscible fluids assisted by a Rayleigh-Taylor instability was explored in Ref.~\onlinecite{lyubimova2019}. In this case, contrary to the case treated in Ref.~\onlinecite{magaletti_jfm_2013}, the diffusive mass transport must not be be neglected. Hence, while many studies show that the use of the CHNS model can well reproduce the Rayleigh-Taylor instability both in 2D and 3D, the thermodynamic parameters are usually considered as free parameters of the model. The way they are chosen is not always explained and if so, their influence on the results is not quantitatively assessed. The lack of convergences studies makes it difficult to use the CHNS model in practical applications where some accuracy and a good estimate of the error is needed.

In this paper, we show the ability of the CHNS model to capture the 
evolution of Rayleigh-Taylor instabilities, highlighting the role of the thermodynamic parameters on the convergence of the model. In the same spirit as a work performed in a different multiphase flow context\cite{song2019}, we compare our convergence results with the scaling of Ref.~\onlinecite{magaletti_jfm_2013}.
The study is limited to immiscible fluids in the low Atwood number limit where the
difference of the fluid densities is small compared to their sum (the Boussinesq 
approximation is used). We first present the governing equations, 
highlighting the specificity of this diffuse-interface model compared to a sharp interface model. We then describe the 
pseudo-spectral numerical method that we use. We then show the validation of the model on two-dimensional 
Rayleigh-Taylor instabilities in the linear regime, based on the theoretical growth rates, and discuss the impact of the 
thermodynamic parameters on the accuracy of the results. We finally report results on three-dimensional 
Rayleigh-Taylor instabilities in the nonlinear regime in viscous fluids (low Reynolds number / Stokes regime or intermediate Reynolds number configurations: $\Re \sim 10^{-2}$ to $10^2$). We observe 
here the influence of the viscosity and of the wavelength of an initial
single-mode perturbation on the mass transport, and discuss briefly the transition
toward the high Reynolds number regime.


\section{Governing Equations}

In this section, we briefly present the Cahn-Hilliard model considered (in the absence of fluid flow), then detail how it is coupled to the Navier-Stokes equations to 
obtain the CHNS model and finally discuss the considered physical properties. 

\subsection{Cahn-Hilliard model (in the absence of fluid flow)}

The Cahn-Hilliard model\cite{cahn_jchp_1958, cahn_jchp_1965} originally aimed at describing the phase separation of 
two chemical species that separate into two homogeneous phases and was inspired by earlier works of van der Walls 
on phase separation. For this purpose, a concentration field is introduced. Here, the considered system of two immiscible 
fluids is seen as a system of two fluids 1 and 2 with miscibility gap and this concentration field, denoted by $c$, represents the molar 
fraction of fluid 2. In the absence of fluid flow, the concentration field obeys the simple diffusion equation
\begin{equation}
\partial_t c= M \LAP \mu,
\label{eq:difusion}
\end{equation}
where $M$ is the mobility and $\mu$ the diffusion potential (the difference between the chemical potentials of fluids 2 
and 1: $\mu = \mu_2 - \mu_1$), which derives from a free energy functional:
\begin{align}
\mu &= \frac{\delta \mathcal{F}}{\delta c},\label{eq:mu}\\
\mathcal{F} &= \int \left( f_0(c) + \frac{\epsilon}{2} (\GRAD c)^2 \right) dV,\label{eq:F}
\end{align}
where $f_0$ is the free energy by unit volume of an homogeneous system and $\epsilon$ a positive constant. Note that 
a constant molar volume is assumed. In our case, $f_0$ is of the form
\begin{equation}
f_0(c) = Ac^2(1 - c)^2,
\label{eq:f0}
\end{equation}
where $A$ is a positive constant. 
$A$ and $\epsilon$ are model parameters that relate to parameters with more physical meaning, as shown next.
With this system of equations, one can prove that the thermodynamic equilibrium 
solution is a system separated into two domains with concentrations 0 and 1 (the two minima of $f_0$). This is appropriate to describe immiscible 
fluids because it means that the phases are composed of fluid 1 or 2 exclusively. Assuming an infinite domain, the 
concentration profile at thermodynamic equilibrium is of the form\cite{cahn_jchp_1958}
\begin{equation}
c(X) = \frac{1}{2} \left( 1 + \tanh \left( \frac{X}{\wint} \right) \right),
\end{equation}
where $X$ is the position on an axis perpendicular to the interface and such that the interface is at $X=0$ and
\begin{equation}
\wint = \sqrt{\frac{2 \epsilon}{A}}
\label{eq:wint}
\end{equation}
can be defined as the interface thickness. The interfacial free energy of the system is defined by\cite{cahn_jchp_1958}
\begin{equation}
\gamma = \int_0^1 \sqrt{2 \epsilon f_0(c)} dc
\end{equation}
at thermodynamic equilibrium. This quantity corresponds to the surface tension between the fluids. By replacing $f_0$ 
with \refeq{eq:f0} and integrating, we have
\begin{equation}
\gamma = \sqrt{\frac{A \epsilon}{18}}.
\label{eq:gamma}
\end{equation}
One can note that \refeqs{eq:wint}{eq:gamma} can be easily reversed to obtain:
\begin{align}
A &= \frac{6 \gamma}{\wint}, \\
\epsilon &= 3 \wint \gamma.
\end{align}
 
\subsection{Coupled Cahn-Hilliard / Navier-Stokes model}

We now describe the coupled CHNS model that describes the evolution of the concentration, velocity and pressure 
fields in the presence of fluid flow. This model is similar to the ones used in previous work\cite{kendon_jfm_2001, 
celani_jfm_2009, lee_ijnme_2011, nitschke_jfm_2012, lee_cma_2013}. The Cahn-Hilliard equation in \refeq{eq:difusion} 
is modified to take into account the effect of the fluid flow through the use of a convective term. It writes then
\begin{equation}
\pd_t c + \bv \SCAL \GRAD c = M \LAP \mu,
\label{eq:diff}
\end{equation}
where $\bv$ is the velocity.

The Navier-Stokes equations (under the Boussinesq approximation) are then extended with a source term to take into account the capillary force:
\begin{multline}
\pd_t \bv + \DIV (\bv \OCROSS \bv) + \frac{1}{\rho_1} \GRAD p - \DIV ( 2 \nu(c) \GRAD^s \bv ) =\\
- \frac{1}{\rho_1} \left( c - \frac{1}{2} \right) \GRAD \mu + \frac{\rho(c)}{\rho_1} \bg,\label{eq:momentum}
\end{multline}
\begin{equation}
\DIV \bv = 0,
\label{eq:incomp}
\end{equation}
where $p$ is the pressure, $\nu$ the kinematic viscosity, $\GRAD^s \bv$ the shear rate tensor defined by
\begin{equation}
\GRAD^s \bv = \frac{1}{2} \left( \GRAD \bv + (\GRAD \bv)^T \right),
\end{equation}
$\rho_i$ the density of fluid $i$ and $\bg$ the gravity. The term associated to the diffusion potential in 
\refeq{eq:momentum} represents the capillary force and derives from a stress tensor analogous to that of 
Korteweg\cite{anderson_arfm_1998}. The dynamic viscosity is assumed to depend linearly on $c$ so that
\begin{equation}
\nu(c) = \frac{\eta_1 + (\eta_2 - \eta_1) c}{\rho_1},
\end{equation}
where $\eta_i$ is the dynamic viscosity of fluid $i$. The density is also assumed to depend linearly on $c$:
\begin{equation}
\rho(c) = \rho_1 + (\rho_2 - \rho_1) c.
\label{eq:density}
\end{equation}

\subsection{Physical properties}

In the whole work, the following densities and surface tension are considered: $\rho_1 = 1000~\densUnit$, 
$\rho_2 = 1010~\densUnit$ and $\gamma = 0.05~\gammaUnit$. These properties are close to those of water. The 
viscosities are varied and are taken equal to 0 when considering perfect fluids. The standard acceleration of gravity is $g = 9.80665~\accUnit$.

Regarding the interface thickness and the mobility, we use various values detailed later. We nevertheless find it 
interesting to briefly discuss here the way these parameters, that are not directly related to quantities usually used to 
describe fluid flows, should be chosen. Let us first discuss the choice of the interface thickness. In actual systems, the 
interface thickness is of the order of a nanometer. From a numerical point of view, simultaneously resolving scales 
ranging from nanometers to meters is clearly out of reach. We can nevertheless expect that the difference between a 
sharp-interface model and a diffuse-interface model will vanish when $w_{int}/L$, where $L$ is the characteristic length 
scale of the problem, tends to zero\cite{magaletti_jfm_2013}. Therefore, $w_{int}$ should be taken small compared to 
the characteristic length scale of the problem.

As also explained in Ref.~\onlinecite{magaletti_jfm_2013}, the choice of the value of the mobility, which can also be considered as a kinetic parameter, involves to take into account two effects 
of the diffusion. At the scale of the interface between the phases, it restores
the thermodynamic equilibrium profile that has 
been deformed by advection. This effect is needed since the capillary force is function of 
the concentration profile and the
actual surface tension is retrieved only if the 
concentration profile of the interface is the equilibrium profile.
Otherwise, the surface tension  is overestimated. On the macroscopic scale, the 
diffusion process tends to smoothen the interface through the diffusive transport of matter
from high curvature regions to 
low curvature regions. This induces a motion of the interface whose velocity
is function of the mobility. This velocity 
in actual systems is negligible compared to the fluid flows and $M$ must therefore be chosen accordingly in the 
model. As a result, the choice of $M$ is a compromise between a sufficiently fast diffusive process (compared to fluid 
flow), that ensures that the interface profile stays the equilibrium profile, and a sufficiently slow one, that avoids the 
appearance of a significant interface motion due to diffusion and not fluid
flow. We find it important to mention that to satisfy these two
requirements  a sufficient scale separation between the
interface thickness and the macroscopic length scale is needed.  

\section{Numerical method}

The model equations are solved using a pseudo-spectral method that has already been used in 
Refs.~\onlinecite{Henry2018,Henry2019}. It relies on the use of the Fourier transform in a periodic domain along directions $x$, $y$ 
and $z$. The solution fields are assumed periodic of periods $N_x \dx = L_x$ in the $x$ direction, $N_y \dx = L_y$ in 
the $y$ direction and $N_z \dx = L_z$ in the $z$ direction, where $\dx > 0$ is the mesh size and $N_x \in \N$ (resp. 
$N_y$/$N_z \in \N$) is the number of Fourier modes used in the $x$ (resp. $y$/$z$) direction. The solving consists in 
computing the $N_x \times N_y \times N_z$ Fourier coefficients associated to the solution fields at every time 
$t^n = n \dt$, where $n \ge 0$ and $\dt > 0$ is the time step. The nonlinear terms of the governing equations are treated 
explicitly while the linear terms are treated implicitly using a simple first order Euler time stepping scheme. The 
incompressibility condition is imposed through a standard Helmholtz projection method. 
Owing to the explicit treatment of the nonlinear terms, the time step is limited by the Courant-Friedrich-Lewy condition.

We use sufficiently high domains so that the interfaces at the top and bottom boundaries are remote from the unstable interface and do not affect the development of the instability, up to the accuracy of our computations.

\section{Linear Analysis in Two Dimensions}

As mentioned above, a better understanding of the quality of the approximation of the sharp-interface multiphase flow 
equations by the CHNS model is needed. To this purpose, we consider the onset of the Rayleigh-Taylor instability. In 
this case, the dispersion relation of the instability is known analytically in the case of perfect fluids and is known with 
good accuracy in the case of viscous fluids. In this section, we first describe the problem setup, focusing on the initial 
condition and on the method used to extract the growth rate. We then present numerical results in the case of perfect 
fluids, which allows us to discuss the convergence of the model regarding the thermodynamic parameters. We finally 
present numerical results obtained in the case of viscous fluids.

\subsection{Problem setup}

The dispersion relation is established in a system that is a $[0,L_x] \times [0,L_z]$ domain where the gravity field is oriented 
along the $z$ axis: $\bg = - g \be_z$, $g > 0$. An initial perturbation of the horizontal interface 
(where the heavy fluid is above the lighter one)
\begin{equation}
\Delta h (t=0) = \zeta_0 \cos(k x) ,~~~~~ k > 0,
\end{equation}
will grow exponentially in the earliest times:
\begin{equation}
\Delta h (t) = \zeta(t) \cos(k x),~~~~~ \zeta(t) = \zeta_0 e^{\alpha t},
\end{equation}
where $\alpha$ is the growth rate given by
\begin{equation}
\alpha = \sqrt{\frac{\rho_2 - \rho_1}{2\rho_1} g k - \frac{\gamma}{2\rho_1}k^3}
\label{eq:growth_rate}
\end{equation}
in the case of perfect fluids, or obtained by solving an equation detailed in Appendix~\ref{app:eta_visc} in the case of 
viscous fluids, under the condition that the wavenumber $k$ is smaller than a critical wavenumber $k_c$ given by
\begin{equation}
k_c = \sqrt{\frac{(\rho_2 - \rho_1)g}{\gamma}}.
\label{eq:kc}
\end{equation}
The formula in \refeq{eq:growth_rate} is adapted from the classical one (see Ref.~\onlinecite[p.~435, Eq.~(51)]{chandrasekhar_clarendon_1961} for instance), where the denominators $2\rho_1$ are 
replaced by $\rho_1 + \rho_2$, owing to the use of the Boussinesq approximation. With the considered physical parameters, we have $k_c \approx 44~\text{m}^{-1}$.

The aim of this study is to compute the growth rates of single-mode perturbations with various wavelengths and to confront them with the 
theoretical growth rates. We therefore use as an initial condition what we consider a good approximation of a single 
mode (transposed in the framework of a diffuse-interface model). In this test, we only consider the unstable wavenumbers
\begin{equation}
k_0 = \frac{2I_0\pi}{L} ,~~~~~ 1 \le I_0 \le E \left( \frac{k_c L}{2\pi} \right),
\label{eq:k0}
\end{equation}
where $L = 1$~m is the maximum considered wavelength. The horizontal domain length is set to one wavelength of the perturbation: 
$L_x = \lambda_0 = 2\pi/k_0 = L/I_0$. The vertical dimension is $L_z = 2L_x$. We initialize the concentration with
\begin{equation}
c_0(x,z) = 
\left\{
\begin{aligned}
&\frac{1}{2} \left( 1 - \tanh \left( \frac{z}{\wint} \right) \right) \\
&~~~~~\text{if } 0 \le z < L_z/4, \\
&\frac{1}{2} \left( 1 + \tanh \left( \frac{z - \frac{L_z}{2} - \zeta_0 \cos(k_0 x)}{\wint} \right) \right) \\
&~~~~~\text{if } L_z/4 \le z < 3L_z/4, \\
&\frac{1}{2} \left( 1 - \tanh \left( \frac{z - L_z}{\wint} \right) \right) \\
&~~~~~\text{if } 3L_z/4 \le z \le L_z, \\
\end{aligned}
\right.
\end{equation}
where $\zeta_0 = 0.5 \times 10^{-4} \lambda_0$ introduces a very small perturbation, and the velocity with
\begin{equation}
\bv_0(x,z) = 
\left\{
\begin{aligned}
&\alpha \zeta_0 e^{k_0 z} (- \sin(k_0 x) \be_x + \cos(k_0 x) \be_z) \\
&~~~~~\text{if } 0 \le z < L_z/2, \\
&\alpha \zeta_0 e^{-k_0 z} (\sin(k_0 x) \be_x + \cos(k_0 x) \be_z) \\
&~~~~~\text{if } L_z/2 \le z \le L_z, \\
\end{aligned}
\right.
\end{equation}
following the analytical solution. 

We then monitor the rise, with respect to $n \ge 0$, of the kinetic energy per unit volume carried by every mode $I_0$,
\begin{equation}
E_{k,0}(t^n) \approx \frac{1}{2} \rho_1 \sum_{K = -N_z/2}^{N_z/2} \left( |v_{x,I_0 0 K}^n|^2 + |v_{z,I_0 0 K}^n|^2 \right),
\label{eq:Ek}
\end{equation}
where $v_{x,I_0 0 K}^n$ and $v_{z,I_0 0 K}^n$ are the $(I_0,0,K)$ Fourier coefficients at $t^n$ of the velocity 
components $v_x$ and $v_z$, respectively. $E_{k,0}(t)$ should grow as $e^{2\alpha t}$ owing to the power 2 over the 
velocity coefficients. In order to compute the growth rate we consider the evolution of this quantity between an initial 
state where the amplitude of the perturbation is $2\zeta_0 = \lambda_0/10^4$ and a final state where it is 
$2\zeta(t_{1/10}) \approx  \lambda_0/10$, \ie for which the nonlinear effects are still negligible. Assuming that the 
analytical prediction of the velocity holds, this final state corresponds to a time
\begin{equation}
t_{1/10} = \frac{1}{\alpha} \ln\left( \frac{\lambda_0}{20 \zeta_0} \right).
\end{equation}
Hence, we can follow the growth of the perturbation amplitude over three orders of magnitude in the linear regime.

\subsection{Interface thickness and mobility}

We now detail how the thermodynamic parameters that have not been defined earlier (interface thickness and mobility) are 
determined. The interface thickness should be small compared to the characteristic size of the perturbation in order to 
approach the sharp interface limit. According to the results of the convergence study on this parameter (see \reffig{fig:RTI_I0_1_wint_effect_Pe_cst}) and to keep the computational cost reasonable, here we set $\wint = \lambda_0/128$ (this value is only changed when performing the 
convergence study). We therefore use a Cahn number (ratio of the interface thickness and the characteristic lengthscale)
\begin{equation}
\Cn = \frac{\wint}{\lambda_0} = \frac{1}{128}.
\end{equation}

The convection needs to dominate the diffusion at the macroscopic scale for the convective flow to develop. This 
requirement is expressed by a strong Péclet number defined as
\begin{equation}
\Pe = \frac{v_l \lambda_0 \wint}{M \gamma},
\label{eq:Pe}
\end{equation}
where $v_l$ is the characteristic velocity. 
However, this number must be not too large so that diffusion is dominant at the interface scale and local thermodynamic equilibrium is enforced at any time. Once again based on the analytical velocity, we can estimate 
that the maximum velocity (at $t_{1/10}$) is
\begin{equation}
v_l = \frac{\lambda_0 \alpha}{20}.
\end{equation}
We set $M$ such that $\Pe = 1000$ (this value is only changed when performing a convergence study on 
this parameter for \reffig{fig:RTI_I0_1_Pe_effect_log}).

These choices are justified by the convergence studies presented in \reffig{fig:convergence}. Note that these studies are performed with a finer mesh than the rest due to the use of smaller values of $\wint$.

\subsection{Numerical parameters}

\label{sec:dx_dt}

Since the interface must be resolved, we use a grid spacing such that there are 4 cells in the interface thickness 
($\dx = \wint/4$), which leads to $N_x = 512$ and $N_z = 1024$ (this is only changed when performing the 
convergence study for \reffig{fig:convergence}). $N_y$ has been set to 1 because we consider 
the two-dimensional case. We then define $\dt$ to at least satisfy the condition
\begin{equation}
\frac{v_l \dt}{\dx} \le 0.25.
\end{equation}
Much smaller values are used to reach convergence for the largest $I_0$ cases.

\subsection{Results in the case of non viscous fluids}

\begin{figure}
\begin{subfigure}{0.5\linewidth}
\centering
\includegraphics[width=0.99\textwidth]{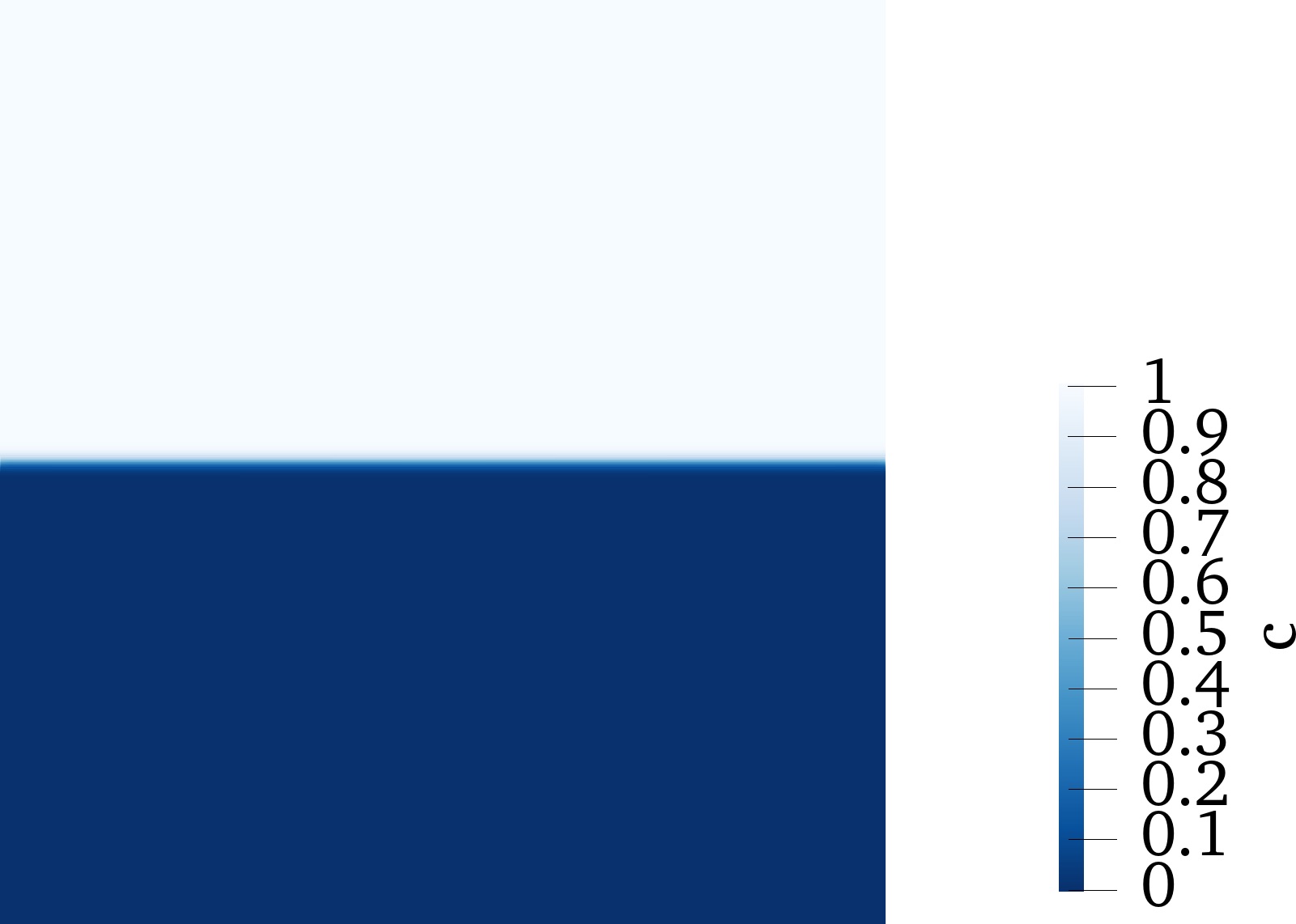}
\caption{$c$ - $t=0$ s}
\end{subfigure}%
\begin{subfigure}{0.5\linewidth}
\centering
\includegraphics[width=0.99\textwidth]{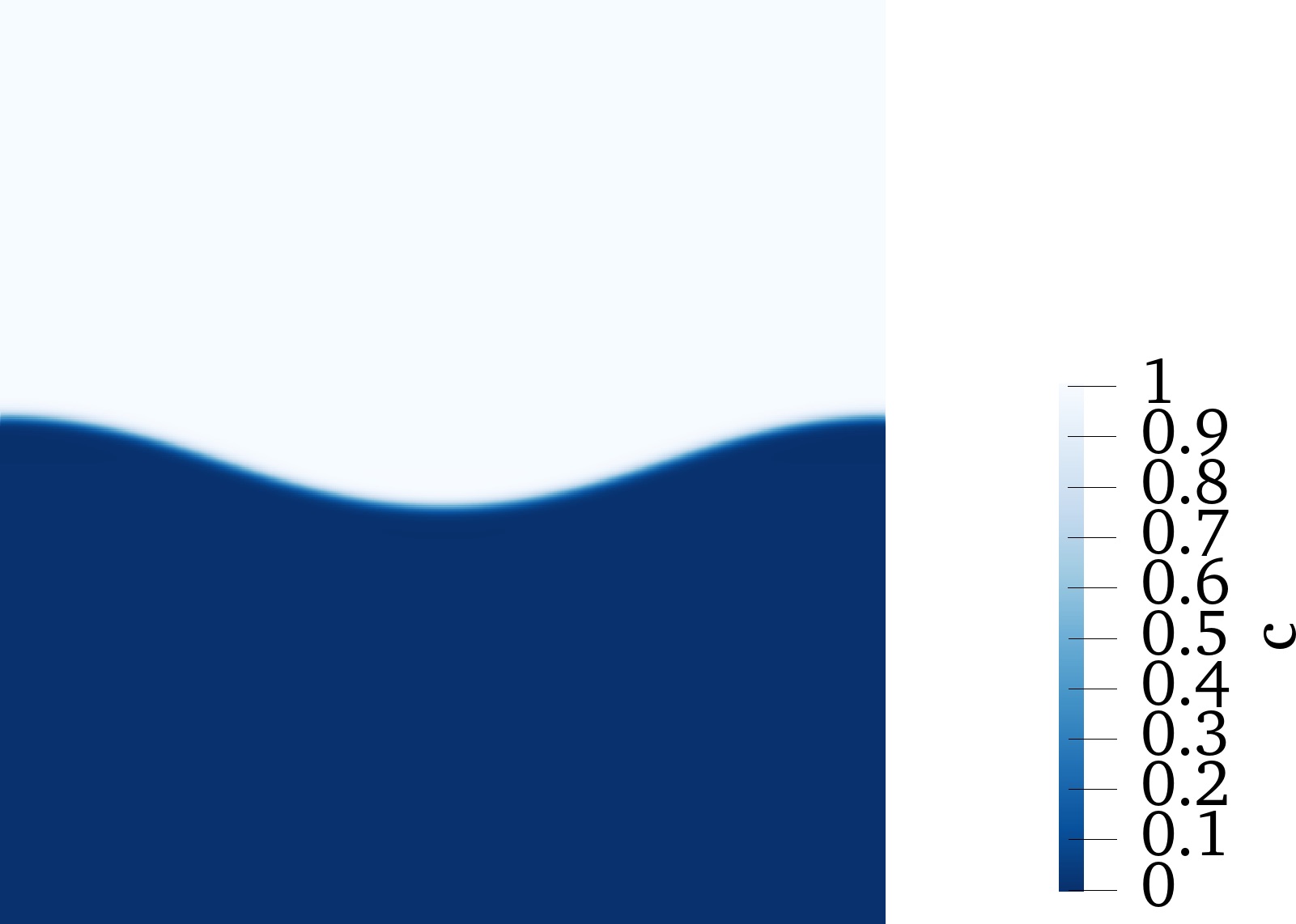}
\caption{$c$ - $t=13$ s}
\end{subfigure}

\begin{subfigure}{0.5\linewidth}
\centering
\includegraphics[width=0.99\textwidth]{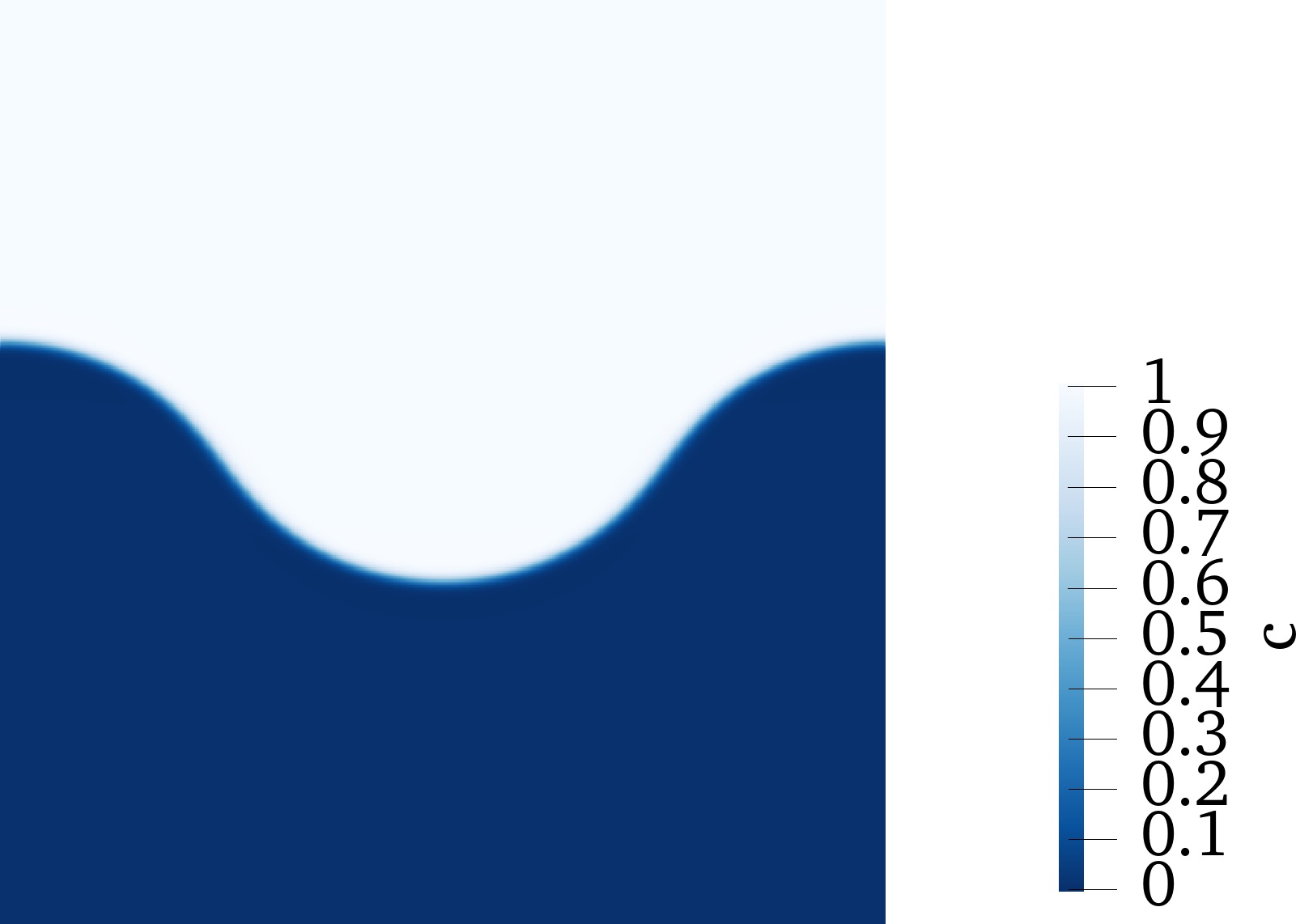}
\caption{$c$ - $t=15$ s}
\end{subfigure}%
\begin{subfigure}{0.5\linewidth}
\centering
\includegraphics[width=0.99\textwidth]{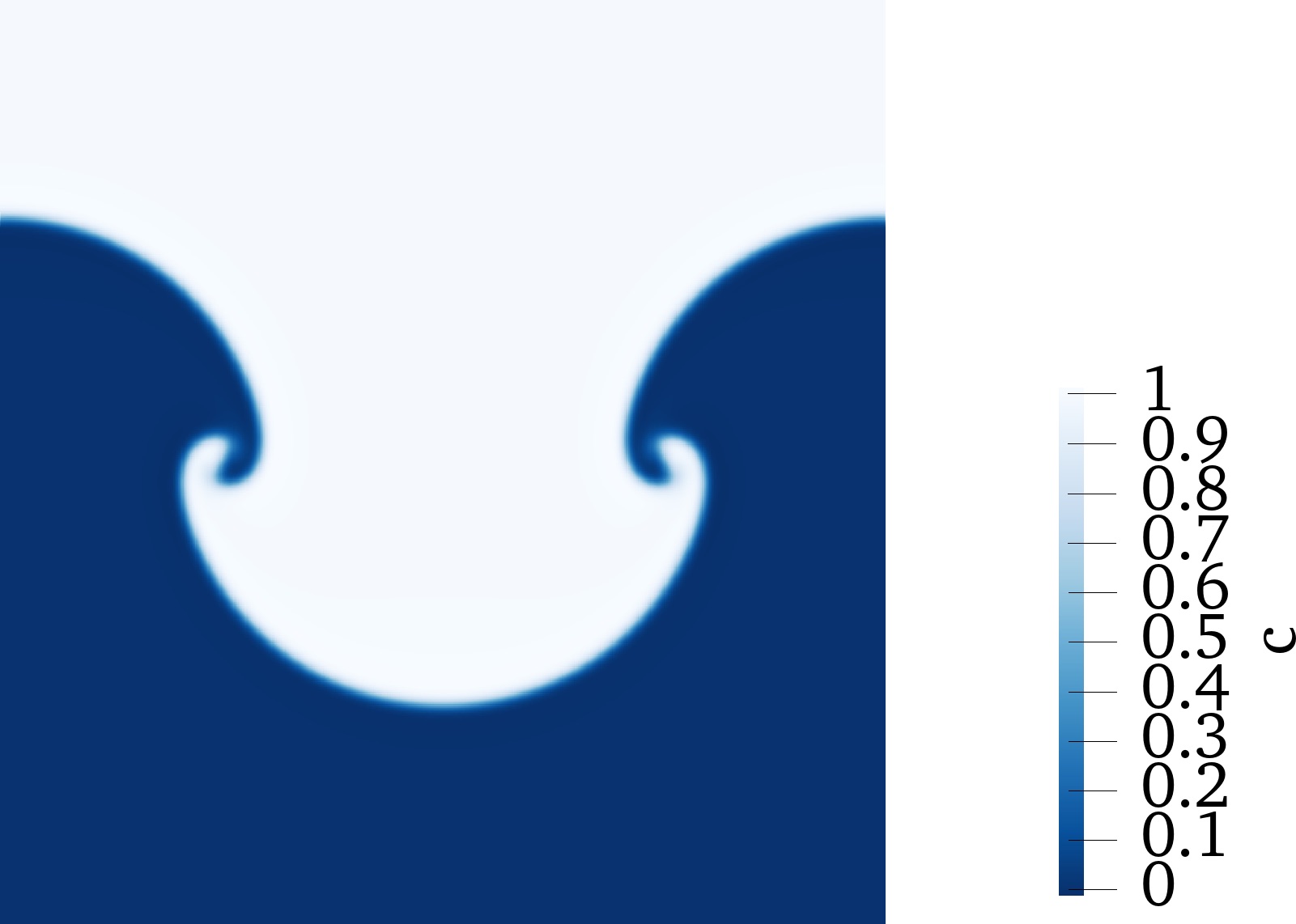}
\caption{$c$ - $t=17$ s}
\end{subfigure}

\begin{subfigure}{0.5\linewidth}
\centering
\includegraphics[width=0.99\textwidth]{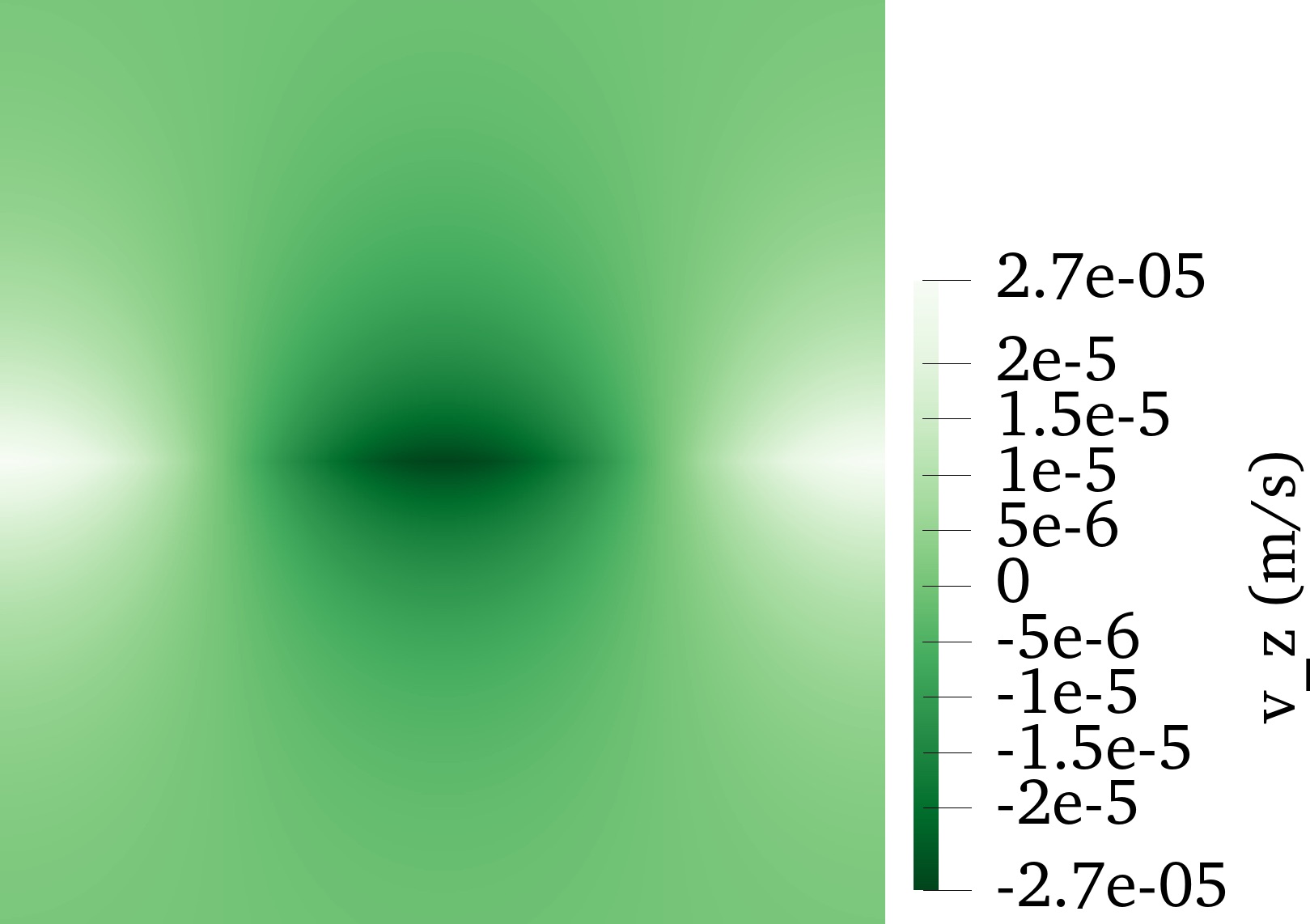}
\caption{$v_z$ - $t=0$ s}
\end{subfigure}%
\begin{subfigure}{0.5\linewidth}
\centering
\includegraphics[width=0.99\textwidth]{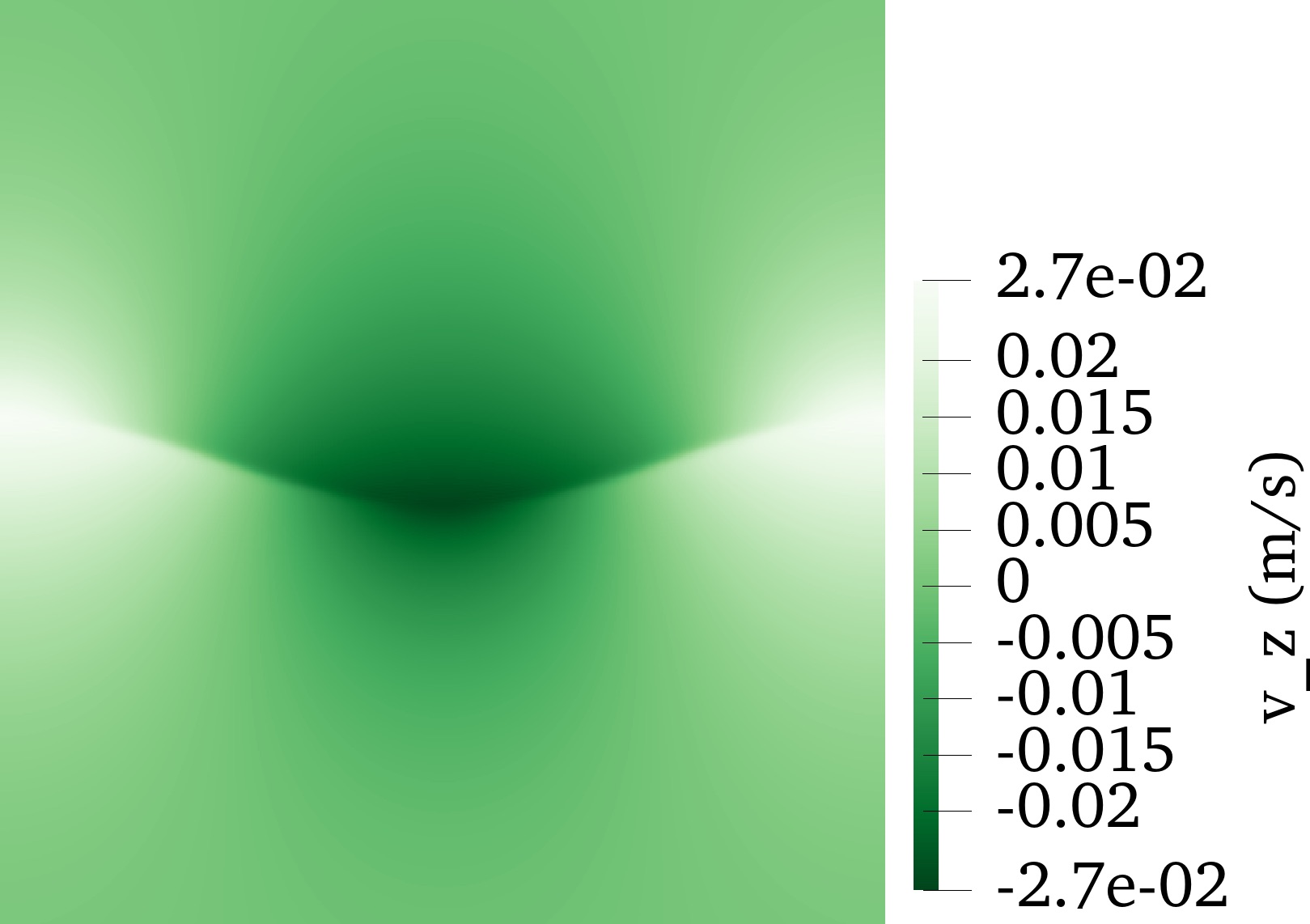}
\caption{$v_z$ - $t=13$ s}
\end{subfigure}

\begin{subfigure}{0.5\linewidth}
\centering
\includegraphics[width=0.99\textwidth]{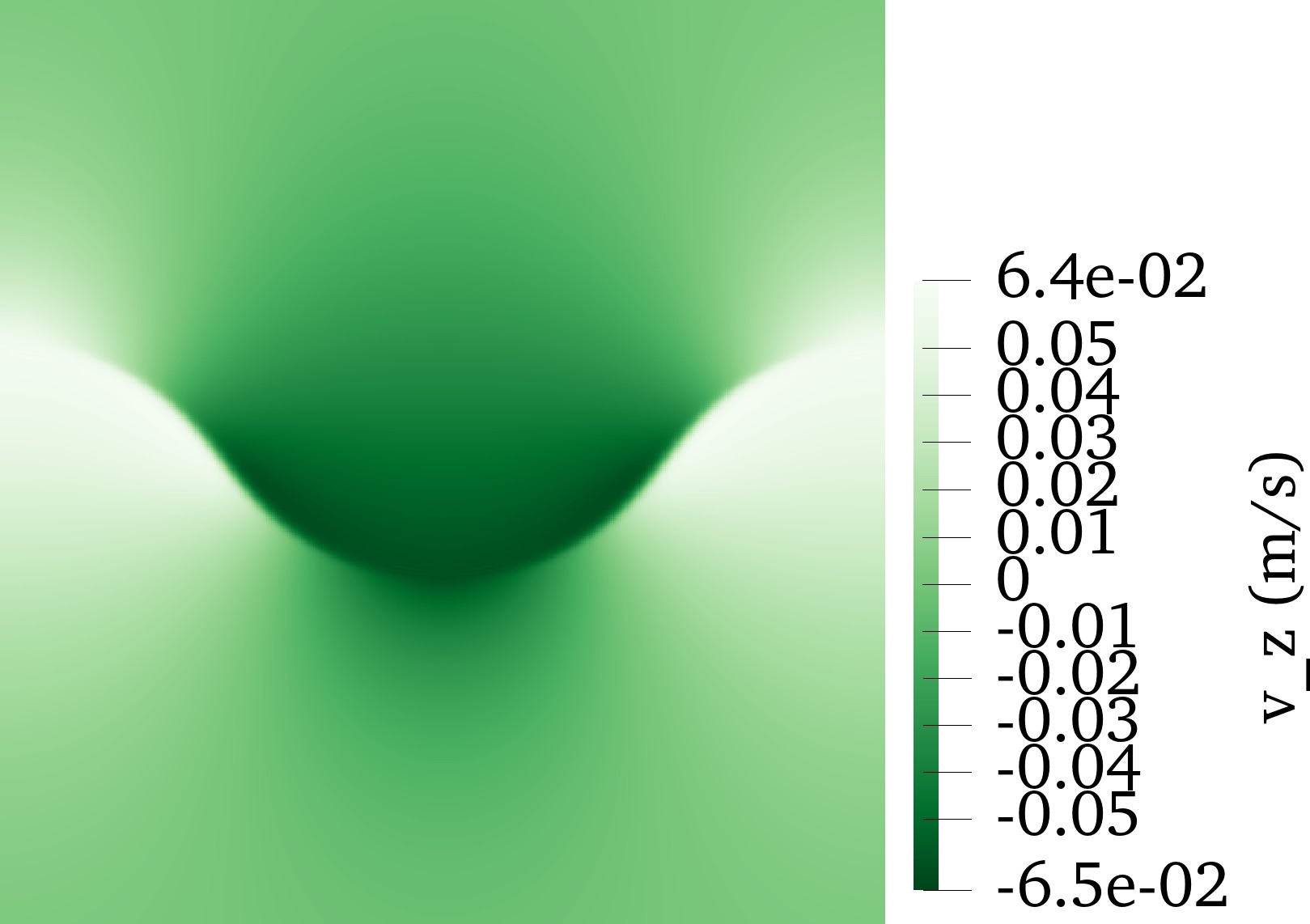}
\caption{$v_z$ - $t=15$ s}
\end{subfigure}%
\begin{subfigure}{0.5\linewidth}
\centering
\includegraphics[width=0.99\textwidth]{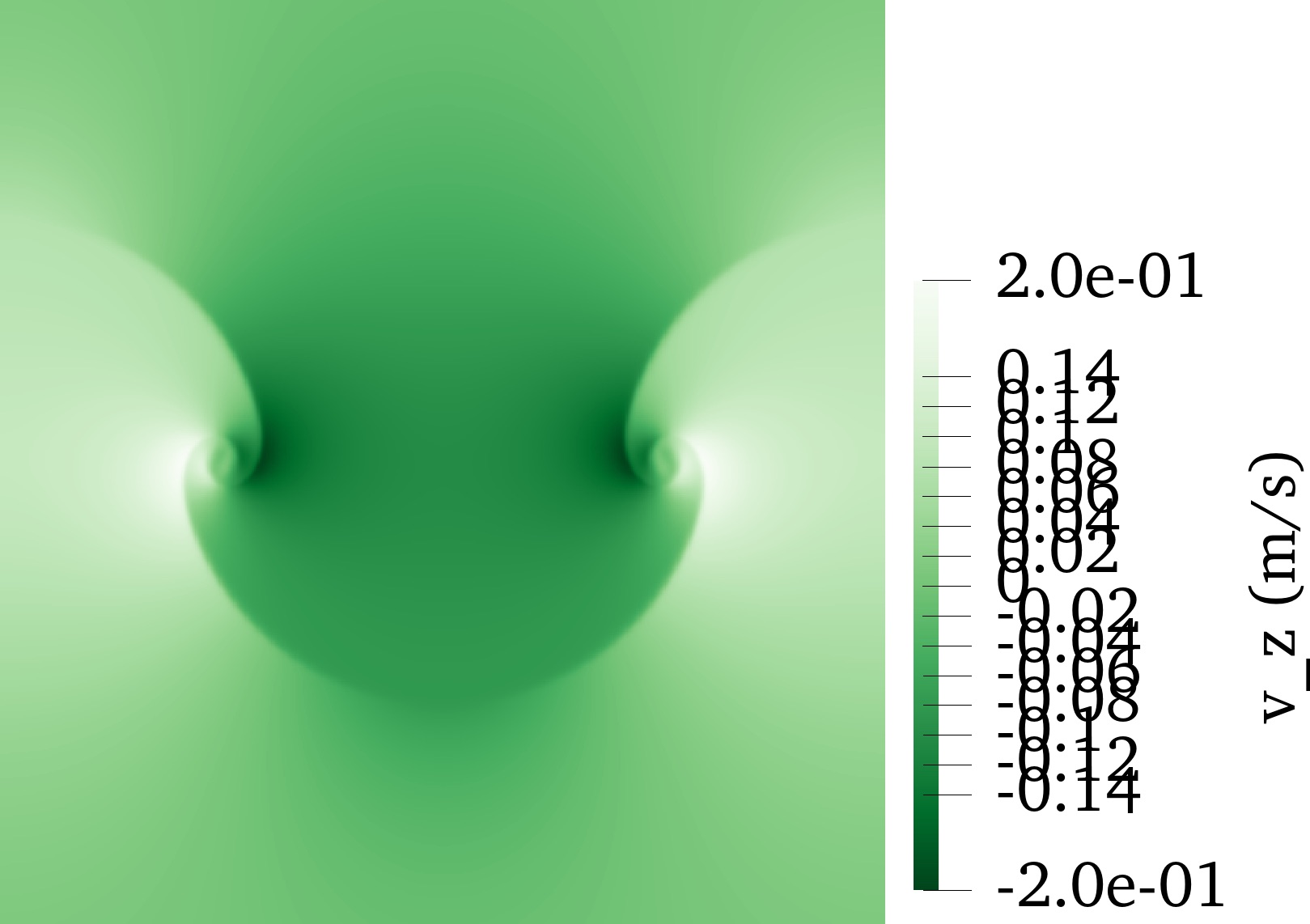}
\caption{$v_z$ - $t=17$ s}
\end{subfigure}
\caption{Concentration and vertical velocity fields during the linear regime ($t < t_{1/10} \approx 12.6$ s) and the nonlinear 
regime of the instability ($t > t_{1/10}$) for $I_0 = 1$.}
\label{fig:sequence}
\end{figure}

First, in order to illustrate quantitatively the development of the instability, we have performed a simulation that goes clearly beyond the linear 
regime ($t > t_{1/10}$) for $I_0 = 1$. In the later moments, since the velocity is much greater than $v_l$, a time step 
much smaller than that obtained by following Section~\ref{sec:dx_dt} is needed. We also use a tiny viscosity 
$\eta = 10^{-3}~\dynViscoUnit$ in this particular case, and beyond the linear regime, to help dissipating the momentum and to avoid too strong 
concentration gradients. During this simulation, the initial perturbation grows exponentially with time and after an initial 
regime where the initial perturbation sine wave is amplified, there is the formation of a spike of heavy fluid falling in 
the light fluid while there is a bubble of light fluid rising in the heavy fluid. This behavior is illustrated in 
\reffig{fig:sequence} where a series of color plots representing the interface between the two phases and the flow 
speed is shown.

\begin{figure}
\centering
\includegraphics[width=\linewidth]{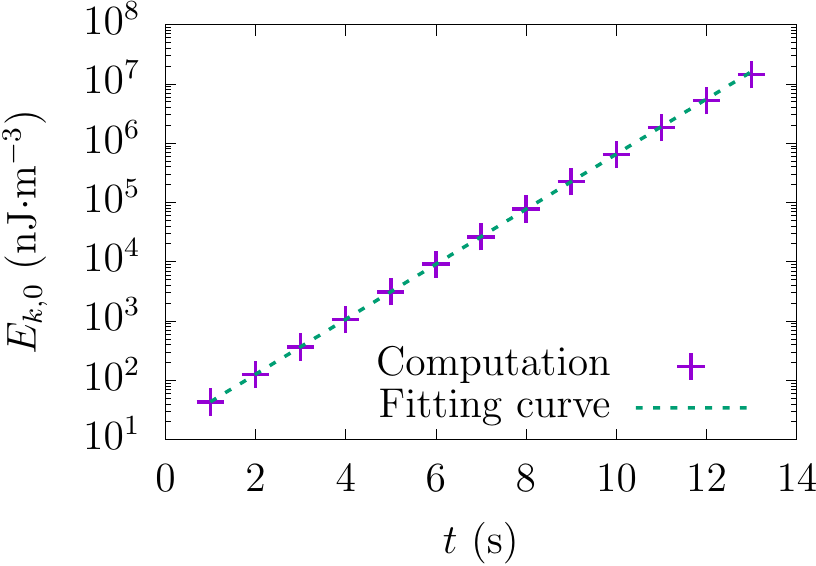}
\caption{Plot of the kinetic energy (using a log scale) carried by mode $I_0$ as a funtion of  time (using a linear scale). The very good match between the data points and the linear fit  indicates that the system is still in the linear regime characterized by an exponential growth. }
\label{fig:RTI_I0_1_Ek_t}
\end{figure}

Hence, the model is able to reproduce qualitatively the phenomenon observed during the Rayleigh-Taylor instability. 
We now turn to a more quantitative analysis where $\eta$ is set to zero. To this purpose, we determine the growth rate of the instability 
(in the linear regime)  by fitting linearly the curve of $\log(E_{k,0}(t))$ between
\begin{equation}
t_{1/1000} = \frac{1}{\alpha} \ln\left( \frac{\lambda_0}{2000 \zeta_0} \right)
\end{equation}
and
\begin{equation}
t_{1/100} = \frac{1}{\alpha} \ln\left( \frac{\lambda_0}{200 \zeta_0} \right),
\end{equation}
the theoretical times needed for the perturbation amplitude to reach one thousandth and one hundredth of the 
wavelength, respectively. The estimated growth rate is the half of the slope of the fitting curve. A typical evolution of 
$E_{k,0}(t)$ and of the fitting curve (that of the case $I_0 = 1$) is shown in \reffig{fig:RTI_I0_1_Ek_t}, which shows 
that the fit is remarkably good. In this case, the fitting is performed between $t_{1/1000} \approx 4.2$~s and 
$t_{1/100} \approx 8.4$~s. The fact that the linear regime is observed without ambiguity from the start is due to our choice of initial condition where the velocity field is not set to zero but to the analytical solution.

\begin{figure}
\centering
\includegraphics[width=\linewidth]{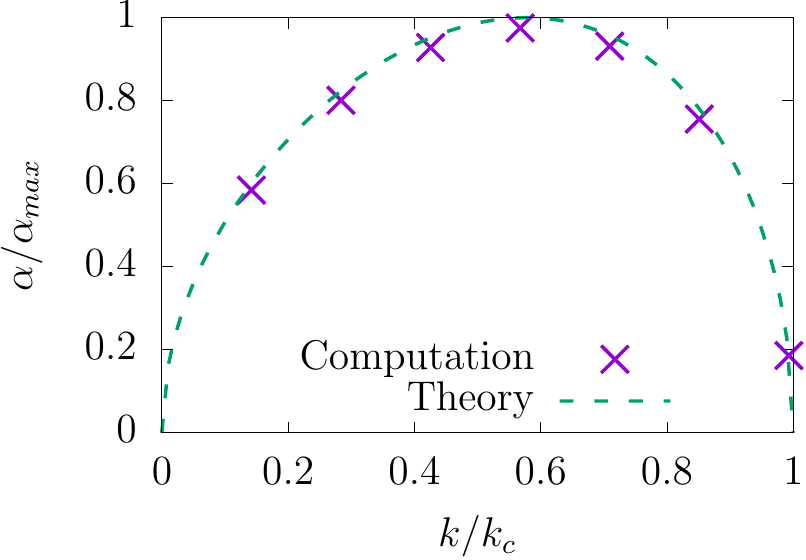}
\caption{Theoretical versus computed growth rates  in the case of perfect fluids. Here, $\alpha_{max} \approx 0.91~\text{s}^{-1}$. }
\label{fig:RTI_disp_diagram}
\end{figure}

From this procedure we are able to extract the growth rate for different values of the wavelength and the obtained 
values are plotted in \reffig{fig:RTI_disp_diagram} together with the theoretical growth rate. The wavenumber is 
normalized by the critical wavenumber, while the growth rate is normalized by the maximum growth rate 
$\alpha_{max}$, which is obtained for the most unstable wavenumber
\begin{equation}
k_{max} = \frac{1}{\sqrt{3}} k_c.
\end{equation}
One can see that the match between the numerical points and the theoretical curve is very good. This indicates that, as 
already discussed in Refs.~\onlinecite{celani_jfm_2009, lee_ijnme_2011} for instance, the CHNS model can quantitatively well 
reproduce the Rayleigh-Taylor instability in the case of perfect fluids.

\begin{figure}
\begin{subfigure}{0.9\linewidth}
\includegraphics[width=\textwidth]{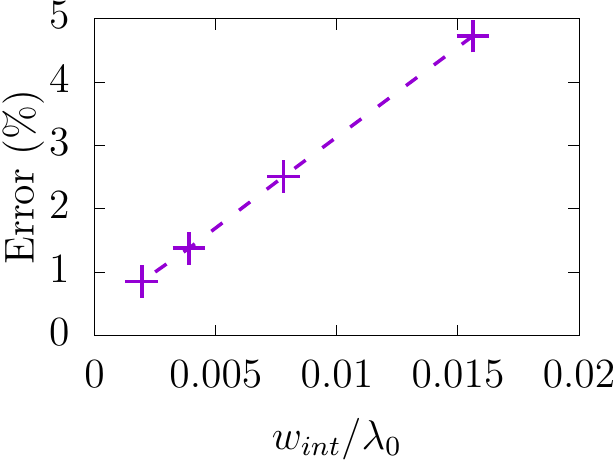}
\caption{Interface thickness effect}
\label{fig:RTI_I0_1_wint_effect_Pe_cst}
\end{subfigure}

\begin{subfigure}{0.9\linewidth}
\includegraphics[width=\textwidth]{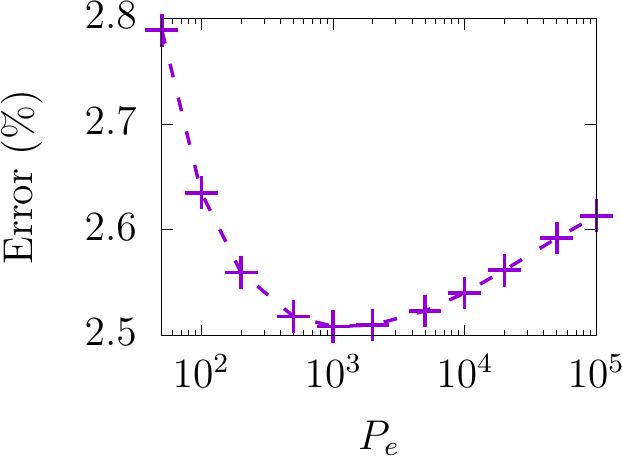}
\caption{Péclet number effect}
\label{fig:RTI_I0_1_Pe_effect_log}
\end{subfigure}

\caption{Influence of the interface thickness and the Péclet number on the error on the growth rate $|(\alpha_{comp}-
\alpha_{th})/\alpha_{th}|$ (case $I_0 = 1$). The cross marks represent computational results. The dashed line is a guide 
for the eye.}
\label{fig:convergence}
\end{figure}

However, there remains a small model-related error. We will now discuss more thoroughly the origin of this error and to 
what extent it can be controlled. The thermodynamic model is fully specified by three parameters: the surface tension, the 
interface thickness and the mobility. The surface tension is determined by the physics of the problem and is therefore 
not a model parameter. The interface thickness and the mobility are model parameters and can be fixed freely (at least 
in the case of the description of the Rayleigh-Taylor instability). Let us first discuss the interface thickness effect. Due to 
the non-zero interface thickness in our model, the numerical solution does not exactly match the analytical one, in 
particular regarding the growth rate. We plot in \reffig{fig:RTI_I0_1_wint_effect_Pe_cst} the relative error on the 
growth rate (case $I_0 = 1$) as a function of the Cahn number. The mobility is such that the Péclet number is 
always $\Pe = 1000$. One can see that the error linearly decreases with this ratio. This therefore confirms that the 
interface thickness must be as small as possible so that the effects of the diffuse interface are limited. However, one can also see that the linear fit does not reach exactly zero when $\Cn$ goes to 0. 
From the analysis\cite{magaletti_jfm_2013}, this is due to the fact that here the Péclet number is taken constant.

Let us now discuss the mobility effect. On the one hand, for a given interface thickness and in the limit of large mobilities, 
the growth of the instability is counterbalanced by the diffusion, which results in an underestimation of the growth rate. 
On the other hand, in the limit of low mobilities, the diffusion cannot restore the proper interface profile, which results in 
an overestimation of the surface tension and therefore in an underestimation of the growth rate. This is illustrated in 
\reffig{fig:RTI_I0_1_Pe_effect_log}, where the relative error on the growth 
rate (case $I_0 = 1$) is plotted as a function of the Péclet number. The 
interface thickness is always $\lambda_0/128$. Recall that the mobility is inversely proportional to the Péclet number. 
One can see that the error decreases with the mobility until a certain value ($\Pe \le 1000$) 
and that it then increases when the mobility is further reduced ($\Pe \ge 1000$). This 
study confirms that the mobility must be small enough to avoid that diffusive transport counterbalances the instability 
mechanism, but also large enough to allow the restoration of the thermodynamic equilibrium profile at the interface by 
diffusive process. These two plots also show that an optimal value of the mobility exists for a given interface thickness 
(here it corresponds to $\Pe = 1000$ approximately).

\begin{figure}
\centering
\includegraphics[width=\linewidth]{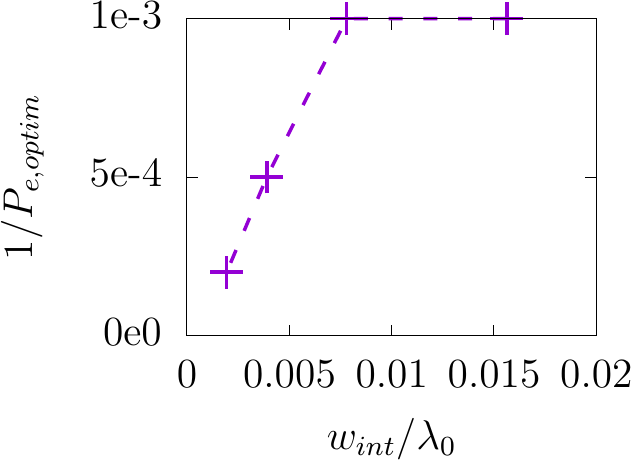}
\caption{Inverse of the optimal Péclet number as a function of the Cahn number.}
\label{fig:RTI_I0_1_Pe_optim}
\end{figure}

The influence of the Péclet number on the accuracy of the result is studied for various values of the interface thickness. Every tested value of $\wint$ presents an optimal value of $\Pe$. We present the inverse of the optimal Péclet number as a function of the Cahn number in \reffig{fig:RTI_I0_1_Pe_optim}. We can see that $1/P_{e,optim}$ is proportional to $\Cn$ when $\Cn$ tends to 0, which is in good agreement with the theoretical work\cite{magaletti_jfm_2013}. In Ref.~\onlinecite{magaletti_jfm_2013}, their optimal dimensionless mobility is predicted to be proportional to the square of the Cahn number but the ratio of the Cahn number and the dimensionless mobility is the Péclet number so it is equivalent.

\subsection{Results in the case of viscous fluids}

\begin{figure}
\includegraphics[width=\linewidth]{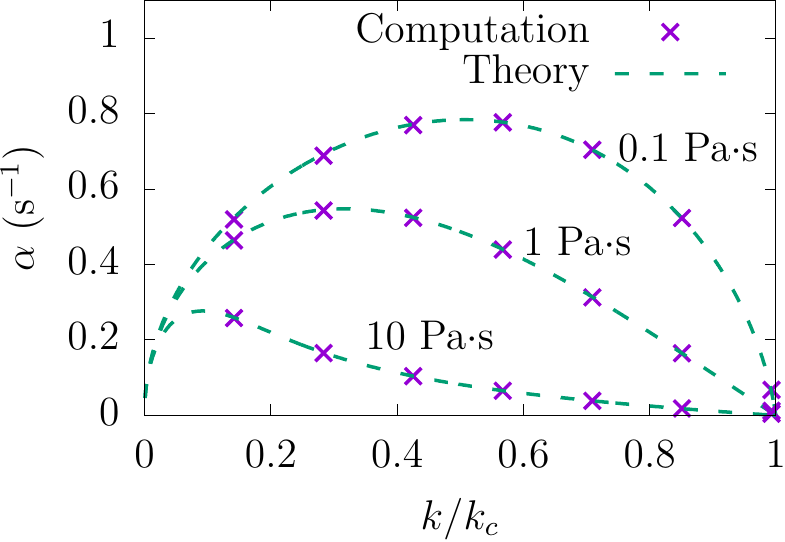}
\caption{Theoretical versus computed growth rates in the case of viscous fluids. Both fluids have the same viscosity.}
\label{fig:RTI_3_visco_disp_diagram}
\end{figure}

We now take advantage of the versatility of the model to study the effects of varying the viscosities of both fluids and 
compare the numerical results with the predicted growth rates presented in Ref.~\onlinecite{menikoff_pof_1977}. First, we consider the case where 
both fluids share the same viscosity, which is such that there is a significant departure from the perfect fluid case 
($\eta_1 = \eta_2 \ge 0.1~\dynViscoUnit$). We plot in \reffig{fig:RTI_3_visco_disp_diagram} the computed growth 
rate as a function of the normalized wavenumber for different values of the viscosity together with the predictions of 
Ref.~\onlinecite{menikoff_pof_1977} (details of the calculation are recalled in Appendix~\ref{app:eta_visc}). As expected (see Ref.~\onlinecite[p.~447]
{chandrasekhar_clarendon_1961}), the wavenumber domain for which the two fluid system is unstable is independent 
of the viscosity (for a given surface tension), which translates into the fact that the curves tend toward a zero growth 
rate for the same value of the wavenumber as in the inviscid case ($k=k_c$). In addition, the agreement between our 
computations and the theoretical predictions is very good. 

\begin{figure}
\includegraphics[width=\linewidth]{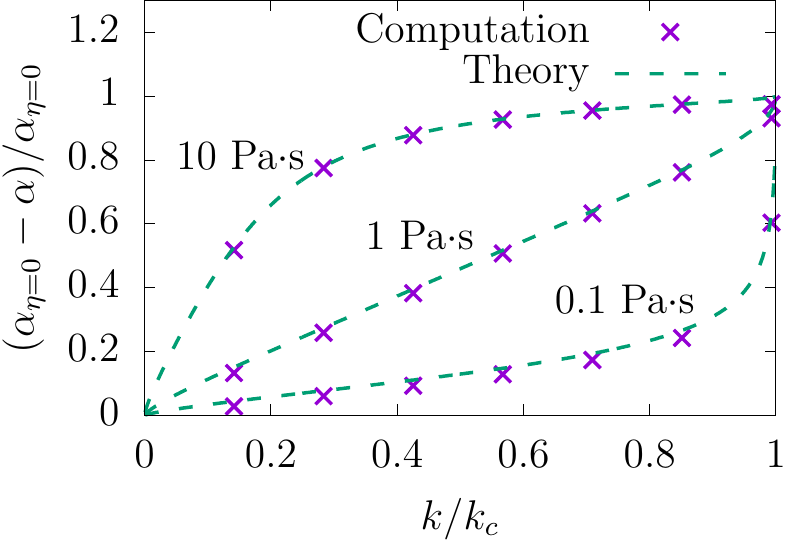}
\caption{Relative difference between the growth rate for perfect fluids $\alpha_{\eta=0}$ and the growth rate for viscous fluids (with the same viscosity).}
\label{fig:RTI_3_visco_disp_diagram_comp_rel}
\end{figure}

We show that the agreement is excellent in 
\reffig{fig:RTI_3_visco_disp_diagram_comp_rel}, where we plot the relative difference between the growth rate for 
perfect fluids and the growth rate for viscous fluids. For small values of the wavenumber, the relative difference is small 
and increases with $k$. This would be expected because of the fact that energy dissipation is higher at high 
wavenumber. The computed slopes close to $k=0$ are in very good agreement with theoretical predictions. At values 
close to $k=k_c$, the theoretical curves have the property that they all converge toward 1 
(while the growth rates converge all toward 0 in $k=k_c$). This behavior is (as can be seen) very well reproduced by 
the numerical points. Similar results were presented in Ref.~\onlinecite{celani_jfm_2009}. However, the comparison was not as compelling since it was limited to the bounds on the growth rates given by Ref.~\onlinecite{menikoff_pof_1977}. The exact growth rates (which can be numerically obtained by solving an equation, see Appendix~\ref{app:eta_visc}) were not considered.

\begin{figure}
\includegraphics[width=\linewidth]{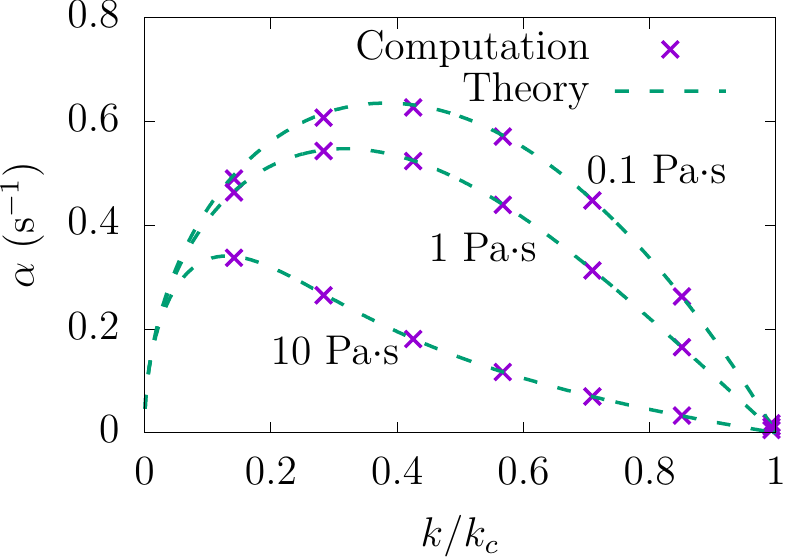}
\caption{Theoretical versus computed growth rates in the case of viscous fluids. The viscosity of the lower fluid stays 
fixed at $\eta_1 = 1~\dynViscoUnit$ while that of the upper fluid varies.}
\label{fig:RTI_4_visco_diff_disp_diagram}
\end{figure}

We now consider that the lower fluid has a fixed viscosity ($\eta_1 = 1~\dynViscoUnit$) while the viscosity of the upper 
fluid is varied and can take a different value ($\eta_2 = 0.1$, 1 or $10~\dynViscoUnit$). The computed versus 
theoretical growth rates for different values of the viscosities are plotted as a function of the normalized wavenumber in 
\reffig{fig:RTI_4_visco_diff_disp_diagram}. Again, the agreement between our computations and the 
theoretical predictions is very good in this case (not treated in Ref.~\onlinecite{celani_jfm_2009}). 

From this study, it is clear that the CHNS model reproduces quantitatively well the 
Rayleigh-Taylor instability in the viscous case as well.

\section{Study of the Nonlinear Regime in Three Dimensions}

We extend the simulations performed in the previous section with three-dimensional simulations of the 
Rayleigh-Taylor instability in the nonlinear regime. 
In particular, we study the effect of the perturbation wavelength and of the viscosity on the 
mass transport in a low-intermediate Reynolds number regime where viscous effects
are significant and for which a steady regime is observed. We also briefly discuss the departure from the steady state when the Reynolds number is increased.
In Ref.~\onlinecite{lee_cma_2013}, three-dimensional simulations in the nonlinear regime were also performed with a similar model (but no surface tension). Here, we take here into account surface tension and perform long-time simulations in high computational domains where the constant bubble velocity and possibly a reacceleration stage can clearly be observed. We believe that these results will supplement existing ones, so that the CHNS model is shown to be able to well reproduce complex multiphase flows.
As in the previous section, we first present the problem setup and then the numerical results.

\subsection{Problem setup}

The system is a $[0,L_x] \times [0,L_y] \times [0,L_z]$ domain where the gravity field is still oriented along the $z$ axis: $\bg = - g \be_z$, $g > 0$. $L_x$ is chosen to be equal to $L_y$ and ranges between $\approx \lambda_c$ and $7 \lambda_c$, 
where
\begin{equation}
\lambda_c = \frac{2\pi}{k_c},
\end{equation}
with $k_c$ defined in \refeq{eq:kc}, is the critical wavelength. 
With the considered physical parameters, we have $\lambda_c \approx 0.14~\lengthUnit$. 
The vertical dimension is taken sufficiently large 
compared to the horizontal dimension to observe the nonlinear regime ($L_z \ge 4L_x$). 
$L_z$ is chosen between $\approx 15 \lambda_c$ and $60 \lambda_c$. 
We initialize the concentration with
\begin{equation}  
c_0(x,y,z) = 
\left\{
\begin{aligned}
&\frac{1}{2} \left( 1 - \tanh \left( \frac{z}{\wint} \right) \right) \\
&~~~~~\text{if } 0 \le z < L_z/4 \\
&\frac{1}{2} \left( 1 + \tanh \left( \frac{z - \frac{L_z}{2} - \zeta_0 p(x,y)}{\wint} \right) \right) \\
&~~~~~\text{if } L_z/4 \le z < 3L_z/4, \\
&\frac{1}{2} \left( 1 - \tanh \left( \frac{z - L_z}{\wint} \right) \right) \\
&~~~~~\text{if } 3L_z/4 \le z \le L_z, \\
\end{aligned}
\right.
\end{equation}
where $\zeta_0$ is taken small compared to the wavelength and
\begin{equation}
p(x,y) = \cos(k_0 x) + \cos(k_0 y),
\label{eq:initperturb}
\end{equation}
with $k_0 = 2\pi/L_x$. Hence, the initial
condition is chosen so that a single wavelength of the initial perturbation is
present in the simulation box ($L_x = L_y = \lambda_0$). The velocity is set to zero initially.

\subsection{Interface thickness and mobility}

In this case, these parameters are the same for every considered wavelength. We use a small value of the interface 
thickness $\wint = L/128$, with $L = 1~\lengthUnit$, compared to the considered wavelengths ($L$ is the maximum considered wavelength).
For $\lambda_0 \approx 1.32 \lambda_c$, the smallest wavelength considered here, we have
$w_{int}/\lambda_0 \approx 0.042$, which would imply an error on the growth rate of
$\approx 10-15 \%$ (extrapolation based on \reffig{fig:RTI_I0_1_wint_effect_Pe_cst}). 

The mobility is set to a small value $M \approx 8.43 \times 10^{-6}~\Munit$ that leads to a Péclet number 
during the nonlinear regime in the range $[5,2000]$ according to \textit{a posteriori}
estimates with a reference fluid velocity being the bubble tip velocity $v_b$ (the velocity of the highest point of the 
interface).

\subsection{Numerical parameters}

We choose here to use 2 cells in the interface thickness ($\dx = \wint/2$), which leads to various values of $N_x$ (and 
$N_y$ which takes the same value) between 48 and 256 depending on $\lambda_0$. $N_z$ goes from 512 to 2048 
depending on $L_z$. The time step is taken small enough to reach convergence ($10^{-3} ~ \timeUnit \le \dt \le 
0.1 ~ \timeUnit$ depending on the viscosity).

\subsection{Results}

Before describing the results, we find it necessary to mention that in our simulations the heavy and light fluids are  
perfectly symmetric in the sense that their motion is symmetric with respect to the mid-plane of the simulation setup (the initial 
position of the interface). We have chosen to focus on the motion of the light fluid. However, one should keep in mind that the perfectly symmetric motion of the heavy fluid is always present.

 With the chosen parameters and the initial condition defined by
 \refeq{eq:initperturb},  the  mode of the initial perturbation is always
 the one dominating during both  the linear 
 and the nonlinear regimes of the instability. In the case of small systems,
 $L_x<2\lambda_c$, this is obvious since the harmonics of the initial
 perturbation correspond to linearly stable modes.
 In the case of larger systems, either the perturbation was growing faster than
 its harmonics (most cases) or the initial contribution of the harmonics was sufficiently
 small to not outgrow the excited mode. As a result, all of what will be
 presented in the following has been obtained through the growth of a single
 mode, or more precisely, of two modes oriented along the $x$ axis and the $y$ axis
 that share the same wavelength.
 
\begin{widetext}
\begin{figure*}
\begin{subfigure}{0.1\linewidth}
\centering
\includegraphics[height=12cm]{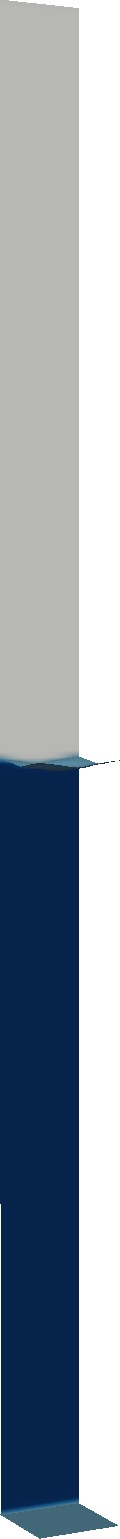}
\caption{0 s}
\end{subfigure}%
\begin{subfigure}{0.1\linewidth}
\centering
\includegraphics[height=12cm]{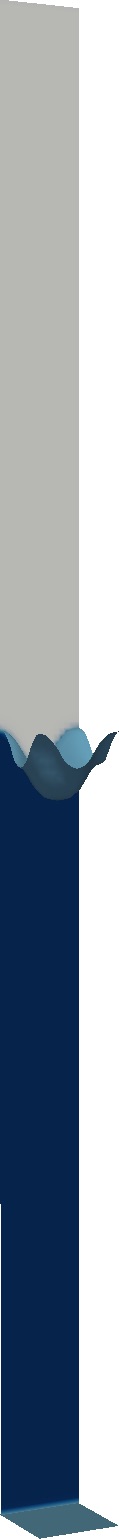}
\caption{8 s}
\end{subfigure}%
\begin{subfigure}{0.1\linewidth}
\centering
\includegraphics[height=12cm]{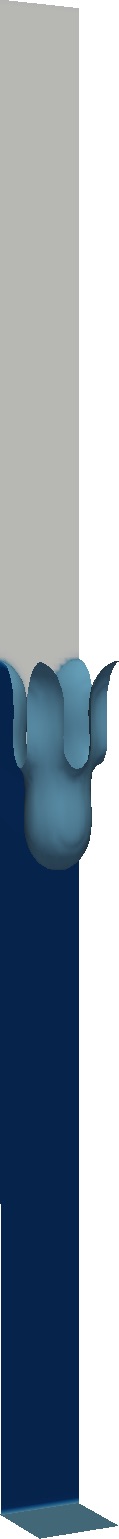}
\caption{16 s}
\end{subfigure}%
\begin{subfigure}{0.1\linewidth}
\centering
\includegraphics[height=12cm]{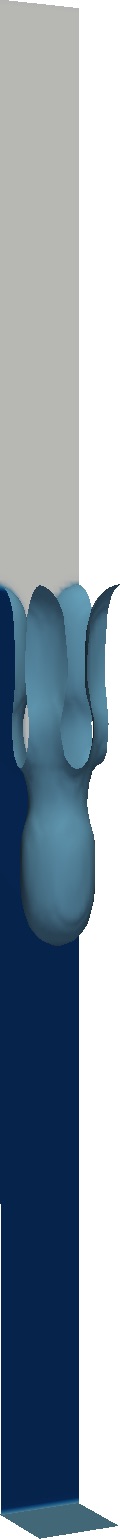}
\caption{24 s}
\end{subfigure}%
\begin{subfigure}{0.1\linewidth}
\centering
\includegraphics[height=12cm]{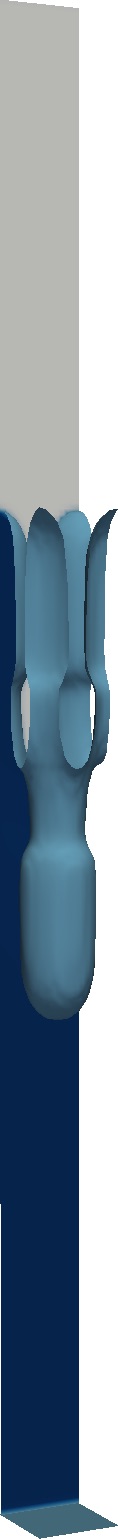}
\caption{32 s}
\end{subfigure}%
\begin{subfigure}{0.1\linewidth}
\centering
\includegraphics[height=12cm]{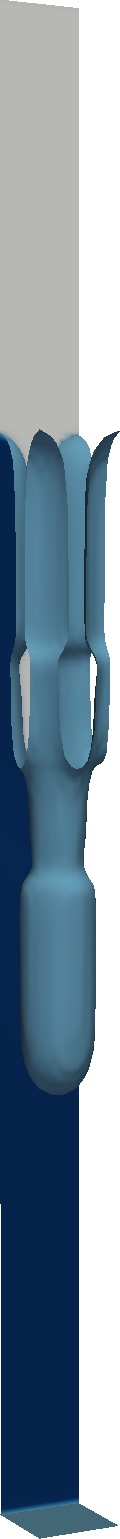}
\caption{40 s}
\end{subfigure}%
\begin{subfigure}{0.1\linewidth}
\centering
\includegraphics[height=12cm]{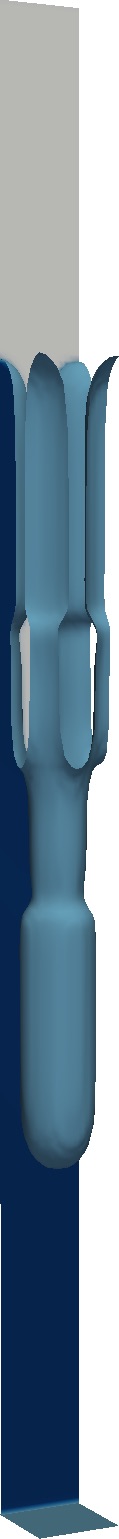}
\caption{48 s}
\end{subfigure}%
\begin{subfigure}{0.1\linewidth}
\centering
\includegraphics[height=12cm]{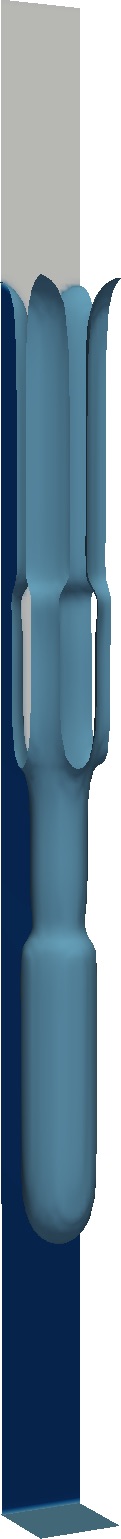}
\caption{56 s}
\end{subfigure}%
\begin{subfigure}{0.1\linewidth}
\centering
\includegraphics[height=12cm]{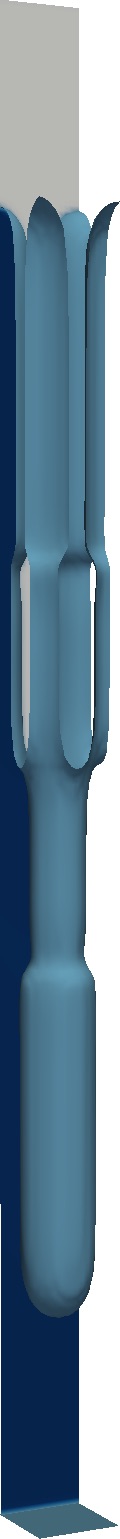}
\caption{64 s}
\end{subfigure}%
\begin{subfigure}{0.1\linewidth}
\centering
\includegraphics[height=12cm]{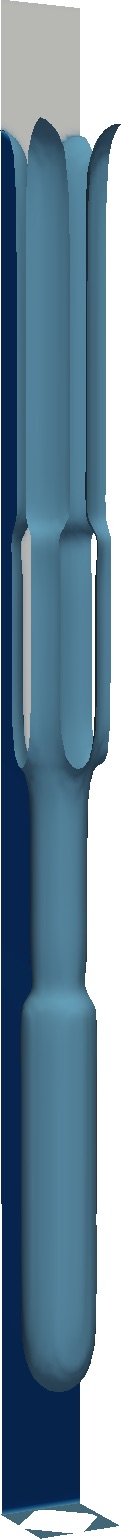}
\caption{72 s}
\end{subfigure}%
\caption{Evolution of the concentration field (dark blue: light fluid, light blue: heavy fluid) and of the interface during the 
evolution of the instability for $\lambda_0 \approx 1.76\lambda_c$ and $\eta = 2~\dynViscoUnit$ ($\Re \approx 3$).}
\label{fig:hh_eta_2_lbda_64_c}
\end{figure*} 
\end{widetext}

 In the whole section, the dynamic viscosities of both fluids are the same and we note it $\eta$.
 The patterns that form during the nonlinear stage are similar to the one
 presented in \reffig{fig:hh_eta_2_lbda_64_c}, obtained for one of the tested cases 
 ($\lambda_0 \approx 1.76\lambda_c$, $\eta = 2~\dynViscoUnit$), that
 is the growth of a bubble of light fluid  followed by a narrower fluid
 filament rising in the heavy fluid. 
 In this case, the Reynolds number, which can be defined as
 \begin{equation}
 \Re = \frac{\rho_1 \max_t (v_b(t)) \lambda_0}{\eta},
 \end{equation}
 is approximately 3.
 From this picture (the snapshot are taken
 at evenly spaced time), the velocity of the bubble tip seems to be constant.
 
\begin{figure}
\centering
\includegraphics[width=\linewidth]{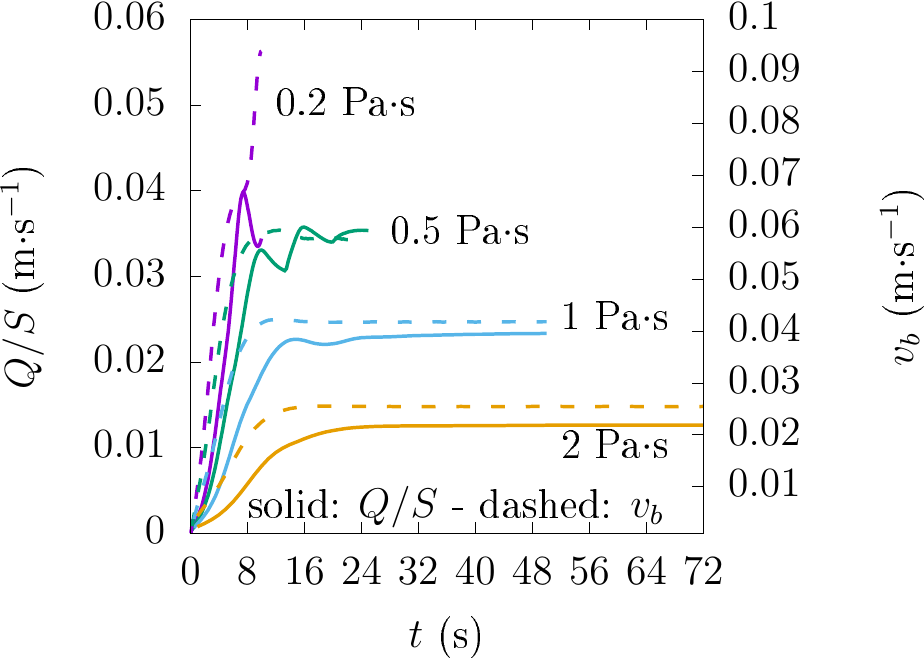}
\caption{$Q/S$ and $v_b$ as a function of time for $\lambda_0 \approx 1.76\lambda_c$ and various $\eta$. For $\eta \ge 0.5~\dynViscoUnit$ ($3 \le \Re \le 30$), after a short transient regime, a steady state is observed. For $\eta = 0.2~\dynViscoUnit$ ($\Re \approx 90$), it is unclear if a steady regime will establish itself because $v_b$ seems to re-increase.}
\label{fig:flowfdet}
\end{figure}

 In order to characterize the mass transport, we use the normalized flow rate $Q/S$ where $Q$ is defined by
 \begin{equation}
 Q = \max_{z \in [0,L_z]} \left( \int_{x=0}^{L_x} \int_{y=0}^{L_y}  (v_z (1-c))(x,y,z) dx \, dy \right) 
 \end{equation}
 and $S = L_x \times L_y$ is the horizontal surface of the simulation domain. 
 $Q$ can be seen as the maximum of the flux of the light (resp. heavy) fluid through the horizontal 
 planes of the domain, counted positively oriented toward the top (resp. the bottom).
 The previous observation is confirmed by the plots in \reffig{fig:flowfdet} where both the normalized flow
 rate and the bubble tip speed are plotted as functions of time. It can be seen
 that, in this case, both quantities converge toward a steady state value after a short transient.

  In \reffig{fig:flowfdet}, one can also see similar plots that were
  obtained for the same wavelength $\lambda_0 \approx 1.76\lambda_c$ and different values
  of the viscosity. From these plots, the convergence toward a steady state is unclear for low values of the
  viscosity. Indeed, when $\eta$ is decreased, one
  can see an alternation of acceleration and deceleration phases on the curve of $Q/S$. This regime is
  similar to damped oscillations and one may expect it to eventually converge
  toward a situation where $Q/S$ is constant with time.
  However, for the lowest value of $\eta$, we observe a re-acceleration stage 
  on the curve of $v_b$, as in Ref.~\onlinecite{ramaprabhu_prE_2006} for instance,
  and a steady regime cannot establish itself. 
  In the following, we will mostly limit ourselves to the configurations where a well defined steady regime is observed.
  We denote by $v_b^{term}$ the terminal bubble velocity.

\begin{figure}
\includegraphics[width=\linewidth]{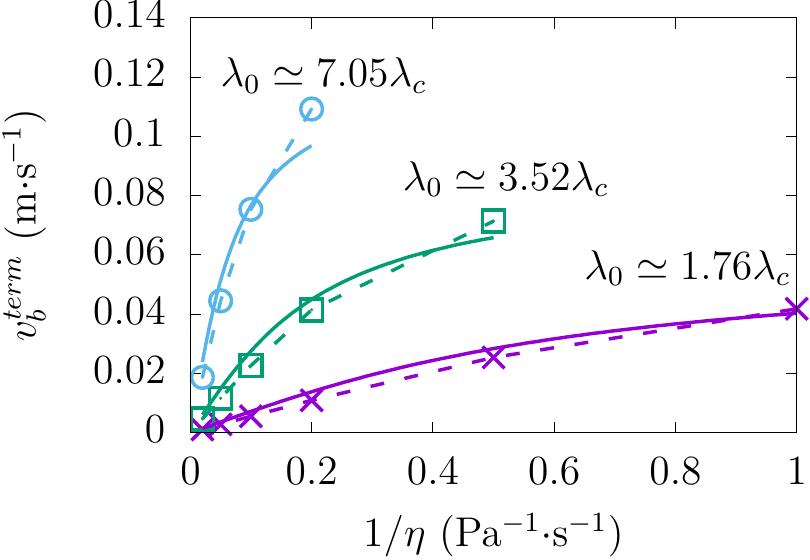}
\caption{Bubble velocity in the steady regime as a function of the viscosity inverse for various values of the wavelength ($0.005 \le \Re \le 20$). The 
symbols represent computational results. The dashed line is a guide for the eye. The solid line follows \refeq{eq:vbt}.
}
\label{fig:vbfdeta}
\end{figure}

To better understand this regime, we first consider the evolution of the terminal bubble velocity as a function of the fluid viscosity. This evolution is presented in \reffig{fig:vbfdeta}, for different values of the wavelength, and the bubble velocity is plotted as a function of $1/\eta$. For high values of $\eta$, \ie small values of  $1/\eta$, the bubble velocity is vanishing like $1/\eta$ as expected in the Stokes regime. As a matter of fact, it is straightforward from the Stokes equation in the permanent regime that the velocity is inversely proportional to $\eta$ in this case. Let us recall that the Stokes velocity (velocity of a sphere falling in a viscous fluid) is proportional to $1/\eta$. And when $\eta$ is further decreased, the bubble velocity increase is sublinear in $1/\eta$. 
In addition to the computational results, we present in \reffig{fig:vbfdeta} the theoretical prediction of Ref.~\onlinecite{sohn_prE_2009} for the terminal velocity of the bubble in the case of a single-mode three-dimensional perturbation. With our notations, the formula translates into
\begin{equation}
v_b^{term} = - k_0 \frac{\eta}{\rho_2} + \sqrt{\frac{2\At}{1 + \At} \frac{g}{k_0} - \frac{3 k_0}{16} \frac{\gamma}{\rho_2} + k_0^2 \left( \frac{\eta}{\rho_2} \right)^2},
\label{eq:vbt}
\end{equation}
where
\begin{equation}
\At = \frac{\rho_2 - \rho_1}{\rho_1 + \rho_2}
\end{equation}
is the Atwood number. Even though it is established for the case of an axisymmetric perturbation, one can see that it follows the computational results reasonably well in the considered range of parameters. Geometrical effects can explain the discrepancy between theory and numerics.
A Taylor expansion applied to \refeq{eq:vbt} shows that, when $k_0$ is fixed and $\eta \to \infty$,
\begin{equation}
v_b^{term} \approx \frac{\rho_2}{2k_0 \eta} \left( \frac{2\At}{1 + \At} \frac{g}{k_0} - \frac{3 k_0}{16} \frac{\gamma}{\rho_2} \right)
\label{eq:vbt_approx}
\end{equation}
so that $v_b^{term} \propto 1/\eta$. This is consistent with the scaling of the velocity with $1/\eta$ observed in the simulations.

\begin{figure}
\begin{subfigure}{0.25\linewidth}
\centering
\includegraphics[scale=0.2]{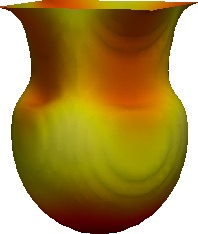}
\caption{10 s}
\end{subfigure}%
\begin{subfigure}{0.25\linewidth}
\centering
\includegraphics[scale=0.2]{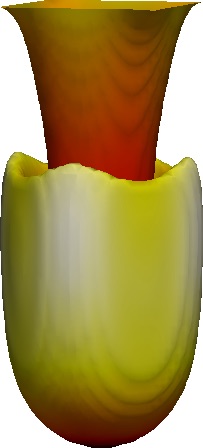}
\caption{15 s}
\end{subfigure}%
\begin{subfigure}{0.5\linewidth}
\centering
\includegraphics[scale=0.2]{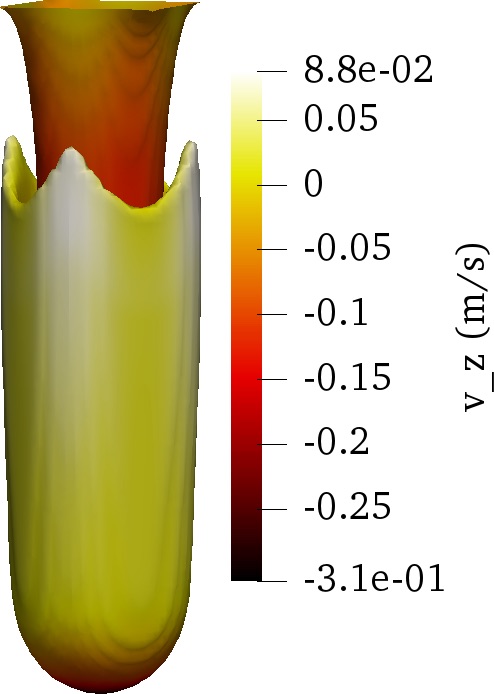}
\caption{20 s}
\end{subfigure}
\caption{Typical pattern formed during the nonlinear stage of the instability at moderate Reynolds number ($\approx 60$). Evolution of the interface, in the bottom half of the domain, colored by the vertical velocity during the evolution of the instability for $\lambda_0 \approx 3.52\lambda_c$ and $\eta = 1~\dynViscoUnit$. One can see that boundary effects are present since the filament presents edges.}
\label{fig:hh_eta_1_lbda_128_vz}
\end{figure}

In the low $\eta$ value cases, the flow is, as expected, reminiscent of what is observed in the case of perfect fluids, as can be seen in \reffig{fig:hh_eta_1_lbda_128_vz}. In this figure, one can see that the flow organizes as follows: the heavy fluid is flowing \textit{rapidly}  downward in the central filament until it reaches the tip of the bubble. It then organizes as an almost  cylindrical  
fluid sheet  that is separated from the filament by a volume of light fluid. This fluid sheet is moving \textit{slowly} upward or is almost at rest. This pattern persists over long time scales. 

\begin{figure}
\includegraphics[width=\linewidth]{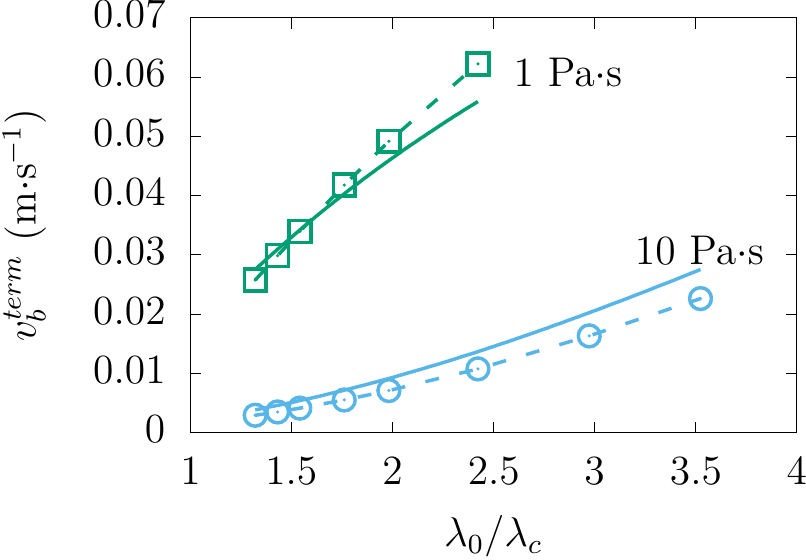}
\caption{Bubble velocity in the steady regime as a function of the wavelength normalized by the critical wavelength for two 
viscosity values ($0.05 \le \Re \le 20$). The symbols represent computational results. The dashed line is a guide for the eye. The solid line follows \refeq{eq:vbt}. 
}
\label{fig:vbfdelbda}
\end{figure}

Finally, for the sake of completeness, we present in \reffig{fig:vbfdelbda}, the terminal bubble velocity as a function of the wavelength normalized by the critical wavelength for two values of the viscosity. In both cases, the bubble velocity increases as a function of $\lambda_0$.
This is due to the fact that energy dissipation is higher at lower wavelength.
Once again, the theoretical prediction of Ref.~\onlinecite{sohn_prE_2009} follows the computational results reasonably well considering that it is not exactly based on the same initial perturbation.

\section{Summary and Discussion}

We have used the CHNS model to simulate the development of Rayleigh-Taylor instabilities in immiscible fluids with a 
low Atwood number and low-intermediate Reynolds numbers. The simulations were performed with a pseudo-spectral code. A first range of simulations was 
performed in two dimensions and at short time to capture the linear regime of the instability, in the case of inviscid or 
viscid fluids. Results show that this kind of diffuse-interface model is perfectly fit to capture this phenomenon, from a 
qualitative but also quantitative point of view. Even though the theory is based on a sharp-interface model, the numerical 
growth rates were very close to the theoretical ones (error of a few percents).

In addition, a careful convergence study has allowed us to elucidate the influence of the thermodynamic parameters on the convergence of the model. The accuracy of the CHNS model is limited by three main effects that are controlled by two parameters of the thermodynamic model: the interface thickness and the mobility. The interface thickness leads to quantitative errors that vanish linearly with the Cahn number $\wint/L$, where $L$ is the characteristic lengthscale of the problem. By using a 1/100 ratio approximately, we obtained errors of less than 5\% on the growth rate of the instability. The effects of the mobility are more complex. The mobility is responsible for keeping the interface profile between the two fluids at thermodynamic equilibrium so that the source term in the flow equation correctly models the action of the surface tension with no overestimation.
Hence, the mobility must be chosen high enough so that diffusion compensates flow effects at the scale of the interface thickness. Since the interface thickness is usually larger than the actual interface thickness in such models, diffusion is likely to be artificially overestimated. Such an overestimation can lead to inaccuracies if the diffusion fluxes lead to an interface motion that is not negligible. Hence, the mobility must be kept small enough to keep this motion negligible. As a result, the choice of the mobility is constrained by a balance between two opposite requirements: diffusion must be efficient at the scale of the interface thickness while it must be negligible at the macroscopic scale. 
In the case of the linear stability analysis, the inverse of the optimal Péclet number (determining the optimal mobility) was found to be proportional to the Cahn number, close to zero, which is in good agreement with former work\cite{magaletti_jfm_2013}.

A second range of simulations was performed in three dimensions and at long time to study the mass transport in the 
nonlinear regime. The study focused on low-intermediate Rayleigh number regimes. Single-mode perturbations with various wavelength were enforced initially. Results show that, when the inertial effects are negligible, a steady regime 
establishes itself. In this steady regime, the bubble tip velocity is constant. Its 
value depends linearly on the inverse of the viscosity and decreases when the wavelength of the perturbation is reduced. 
A reasonable agreement was found with a theoretical formula for axisymmetric perturbations.

  From this study, we believe that we have provided a clear view of the CHNS
  model advantages and drawbacks in cases where a single lengthscale is present. 
  It is simple to implement numerically and can
  handle without any additional cost topological changes of the phases. This
  comes at the expense of the need of resolving two length scales that must be
  separated by orders of magnitude in order to reach a good accuracy.
  It must also be noted that the choice of thermodynamic parameters is far from
  simple and must be made with care, taking into account the characteristic flow
  speed and lengthscale. In multiscale problems, the choice should be more involved. However, departures from the optimal Péclet number do not necessarily affect strongly the error (see \reffig{fig:RTI_I0_1_Pe_effect_log}). In this case, using an intermediate lengthscale of the problem as reference should prove reasonable.
Let us finally remark that, in the case of turbulent flows, the model will not be able to describe the smaller length scales at which the energy dissipation occurs and that this may affect the convergence of the model.

\appendix

\section{Growth rate of the Rayleigh-Taylor instability in viscous fluids}

\label{app:eta_visc}

Introducing a dimensionless wavenumber, growth rate and surface tension,
\begin{multline}
k^* = \left( \frac{\nu_m^2}{\At g} \right)^\frac{1}{3} k
,~~~
\alpha^* = \left( \frac{\nu_m}{\At^2 g^2} \right)^\frac{1}{3} \alpha \\
\text{and}~~~
\gamma^* = \left( \frac{1}{\At g \nu_m^4 (\rho_1 + \rho_2)^3} \right)^\frac{1}{3} \gamma,
\label{eq:visco_alpha_1}
\end{multline}
where $\nu_m$ is a mean kinematic viscosity defined by
\begin{equation}
\nu_m = \frac{\eta_1 + \eta_2}{\rho_1 + \rho_2},
\end{equation}
the authors of Ref.~\onlinecite{menikoff_pof_1977} claim that the ratio $s = \alpha^*/(k^*)^2$ is root of the 
equation
\begin{equation}
s^2 + 2sH(s) = R,
\end{equation}
where
\begin{equation}
H(s) = \frac{2a(s)b(s)}{a(s) + b(s)},
\end{equation}
\begin{multline}
a(s) = C_2 + \sqrt{C_1 + C_1 A_1 s},\\
b(s) = C_1 + \sqrt{C_2 + C_2 A_2 s},
\end{multline}
\begin{equation}
A_i = \frac{\rho_i}{\rho_1 + \rho_2},
~~~~~
C_i = \frac{\eta_i}{\eta_1 + \eta_2},
~~~~~
i=1,2,
\end{equation}
and
\begin{equation}
R = \frac{1 - \gamma^* (k^*)^2}{(k^*)^3}.
\label{eq:visco_alpha_2}
\end{equation}
We compute an estimate of this root by using Newton's method.



%
%

%

\begin{acknowledgments}
This work was supported by the CEA (Commissariat \`a l'\'energie atomique et aux \'energies alternatives) and EDF (Electricit\'e de France). The computations were performed on the clusters of Ecole Polytechnique and GENCI-IDRIS (project 0042B07727).
We thank Mathis Plapp from PMC Laboratory for the valuable discussions.
\end{acknowledgments}

\section*{Data Availability Statement}

The data that support the findings of this study are available from the corresponding author upon reasonable request.

\bibliography{Biblio}

\begin{thebibliography}{31}%
\makeatletter
\providecommand \@ifxundefined [1]{%
 \@ifx{#1\undefined}
}%
\providecommand \@ifnum [1]{%
 \ifnum #1\expandafter \@firstoftwo
 \else \expandafter \@secondoftwo
 \fi
}%
\providecommand \@ifx [1]{%
 \ifx #1\expandafter \@firstoftwo
 \else \expandafter \@secondoftwo
 \fi
}%
\providecommand \natexlab [1]{#1}%
\providecommand \enquote  [1]{``#1''}%
\providecommand \bibnamefont  [1]{#1}%
\providecommand \bibfnamefont [1]{#1}%
\providecommand \citenamefont [1]{#1}%
\providecommand \href@noop [0]{\@secondoftwo}%
\providecommand \href [0]{\begingroup \@sanitize@url \@href}%
\providecommand \@href[1]{\@@startlink{#1}\@@href}%
\providecommand \@@href[1]{\endgroup#1\@@endlink}%
\providecommand \@sanitize@url [0]{\catcode `\\12\catcode `\$12\catcode
  `\&12\catcode `\#12\catcode `\^12\catcode `\_12\catcode `\%12\relax}%
\providecommand \@@startlink[1]{}%
\providecommand \@@endlink[0]{}%
\providecommand \url  [0]{\begingroup\@sanitize@url \@url }%
\providecommand \@url [1]{\endgroup\@href {#1}{\urlprefix }}%
\providecommand \urlprefix  [0]{URL }%
\providecommand \Eprint [0]{\href }%
\providecommand \doibase [0]{http://dx.doi.org/}%
\providecommand \selectlanguage [0]{\@gobble}%
\providecommand \bibinfo  [0]{\@secondoftwo}%
\providecommand \bibfield  [0]{\@secondoftwo}%
\providecommand \translation [1]{[#1]}%
\providecommand \BibitemOpen [0]{}%
\providecommand \bibitemStop [0]{}%
\providecommand \bibitemNoStop [0]{.\EOS\space}%
\providecommand \EOS [0]{\spacefactor3000\relax}%
\providecommand \BibitemShut  [1]{\csname bibitem#1\endcsname}%
\let\auto@bib@innerbib\@empty
\bibitem [{\citenamefont {Sharp}(1984)}]{sharp_physica12D_1984}%
  \BibitemOpen
  \bibfield  {author} {\bibinfo {author} {\bibfnamefont {D.~H.}\ \bibnamefont
  {Sharp}},\ }\bibfield  {title} {\enquote {\bibinfo {title} {{An overview of
  Rayleigh-Taylor instability}},}\ }\href@noop {} {\bibfield  {journal}
  {\bibinfo  {journal} {Physica 12D}\ ,\ \bibinfo {pages} {3--18}} (\bibinfo
  {year} {1984})}\BibitemShut {NoStop}%
\bibitem [{\citenamefont {Zhou}(2017)}]{zhou_physicsReports_2017}%
  \BibitemOpen
  \bibfield  {author} {\bibinfo {author} {\bibfnamefont {Y.}~\bibnamefont
  {Zhou}},\ }\bibfield  {title} {\enquote {\bibinfo {title} {{Rayleigh–Taylor
  and Richtmyer–Meshkov instability induced flow, turbulence, and mixing.
  I}},}\ }\href@noop {} {\bibfield  {journal} {\bibinfo  {journal} {Physics
  Reports}\ }\textbf {\bibinfo {volume} {720-722}},\ \bibinfo {pages} {1--136}
  (\bibinfo {year} {2017})}\BibitemShut {NoStop}%
\bibitem [{\citenamefont {Taylor}(1950)}]{taylor_rsl_1950}%
  \BibitemOpen
  \bibfield  {author} {\bibinfo {author} {\bibfnamefont {G.}~\bibnamefont
  {Taylor}},\ }\bibfield  {title} {\enquote {\bibinfo {title} {{The instability
  of liquid surfaces when accelerated in a direction perpendicular to their
  planes. I}},}\ }\href@noop {} {\bibfield  {journal} {\bibinfo  {journal}
  {Proc. R. Soc. Lond. A}\ }\textbf {\bibinfo {volume} {201}},\ \bibinfo
  {pages} {192--196} (\bibinfo {year} {1950})}\BibitemShut {NoStop}%
\bibitem [{\citenamefont {Chandrasekhar}(1961)}]{chandrasekhar_clarendon_1961}%
  \BibitemOpen
  \bibfield  {author} {\bibinfo {author} {\bibfnamefont {S.}~\bibnamefont
  {Chandrasekhar}},\ }\href@noop {} {\emph {\bibinfo {title} {{Hydrodynamic and
  Hydromagnetic Stability}}}}\ (\bibinfo  {publisher} {Clarendon Press},\
  \bibinfo {year} {1961})\BibitemShut {NoStop}%
\bibitem [{\citenamefont {Menikoff}\ \emph {et~al.}(1977)\citenamefont
  {Menikoff}, \citenamefont {Mjolsness}, \citenamefont {Sharp},\ and\
  \citenamefont {Zemach}}]{menikoff_pof_1977}%
  \BibitemOpen
  \bibfield  {author} {\bibinfo {author} {\bibfnamefont {R.}~\bibnamefont
  {Menikoff}}, \bibinfo {author} {\bibfnamefont {R.~C.}\ \bibnamefont
  {Mjolsness}}, \bibinfo {author} {\bibfnamefont {D.~H.}\ \bibnamefont
  {Sharp}}, \ and\ \bibinfo {author} {\bibfnamefont {C.}~\bibnamefont
  {Zemach}},\ }\bibfield  {title} {\enquote {\bibinfo {title} {{Unstable normal
  mode for Rayleigh–Taylor instability in viscous fluids}},}\ }\href@noop {}
  {\bibfield  {journal} {\bibinfo  {journal} {Phys. Fluids}\ }\textbf {\bibinfo
  {volume} {20}},\ \bibinfo {pages} {2000--2004} (\bibinfo {year}
  {1977})}\BibitemShut {NoStop}%
\bibitem [{\citenamefont {Lewis}\ and\ \citenamefont
  {Taylor}(1950)}]{lewis_rsl_1950}%
  \BibitemOpen
  \bibfield  {author} {\bibinfo {author} {\bibfnamefont {D.~J.}\ \bibnamefont
  {Lewis}}\ and\ \bibinfo {author} {\bibfnamefont {G.}~\bibnamefont {Taylor}},\
  }\bibfield  {title} {\enquote {\bibinfo {title} {{The instability of liquid
  surfaces when accelerated in a direction perpendicular to their planes.
  II}},}\ }\href@noop {} {\bibfield  {journal} {\bibinfo  {journal} {Proc. R.
  Soc. Lond. A}\ }\textbf {\bibinfo {volume} {202}},\ \bibinfo {pages} {81--96}
  (\bibinfo {year} {1950})}\BibitemShut {NoStop}%
\bibitem [{\citenamefont {Waddell}, \citenamefont {Niederhaus},\ and\
  \citenamefont {Jacobs}(2001)}]{waddell_pof_2001}%
  \BibitemOpen
  \bibfield  {author} {\bibinfo {author} {\bibfnamefont {J.~T.}\ \bibnamefont
  {Waddell}}, \bibinfo {author} {\bibfnamefont {C.~E.}\ \bibnamefont
  {Niederhaus}}, \ and\ \bibinfo {author} {\bibfnamefont {J.~W.}\ \bibnamefont
  {Jacobs}},\ }\bibfield  {title} {\enquote {\bibinfo {title} {{Experimental
  study of Rayleigh–Taylor instability: Low Atwood number liquid systems with
  single-mode initial perturbations}},}\ }\href@noop {} {\bibfield  {journal}
  {\bibinfo  {journal} {Phys. Fluids}\ }\textbf {\bibinfo {volume} {13}},\
  \bibinfo {pages} {1263--1273} (\bibinfo {year} {2001})}\BibitemShut {NoStop}%
\bibitem [{\citenamefont {Wilkinson}\ and\ \citenamefont
  {Jacobs}(2007)}]{wilkinson_pof_2007}%
  \BibitemOpen
  \bibfield  {author} {\bibinfo {author} {\bibfnamefont {J.~P.}\ \bibnamefont
  {Wilkinson}}\ and\ \bibinfo {author} {\bibfnamefont {J.~W.}\ \bibnamefont
  {Jacobs}},\ }\bibfield  {title} {\enquote {\bibinfo {title} {{Experimental
  study of the single-mode three-dimensional Rayleigh-Taylor instability}},}\
  }\href@noop {} {\bibfield  {journal} {\bibinfo  {journal} {Phys. Fluids}\
  }\textbf {\bibinfo {volume} {19}},\ \bibinfo {pages} {124102} (\bibinfo
  {year} {2007})}\BibitemShut {NoStop}%
\bibitem [{\citenamefont {Oron}\ \emph {et~al.}(2001)\citenamefont {Oron},
  \citenamefont {Arazi}, \citenamefont {Kartoon}, \citenamefont {Rikanati},
  \citenamefont {Alon},\ and\ \citenamefont {Shvarts}}]{oron_pop_2001}%
  \BibitemOpen
  \bibfield  {author} {\bibinfo {author} {\bibfnamefont {D.}~\bibnamefont
  {Oron}}, \bibinfo {author} {\bibfnamefont {L.}~\bibnamefont {Arazi}},
  \bibinfo {author} {\bibfnamefont {D.}~\bibnamefont {Kartoon}}, \bibinfo
  {author} {\bibfnamefont {A.}~\bibnamefont {Rikanati}}, \bibinfo {author}
  {\bibfnamefont {U.}~\bibnamefont {Alon}}, \ and\ \bibinfo {author}
  {\bibfnamefont {D.}~\bibnamefont {Shvarts}},\ }\bibfield  {title} {\enquote
  {\bibinfo {title} {{Dimensionality dependence of the Rayleigh–Taylor and
  Richtmyer–Meshkov instability late-time scaling laws}},}\ }\href@noop {}
  {\bibfield  {journal} {\bibinfo  {journal} {Phys. Plasmas}\ }\textbf
  {\bibinfo {volume} {8}},\ \bibinfo {pages} {2883--2889} (\bibinfo {year}
  {2001})}\BibitemShut {NoStop}%
\bibitem [{\citenamefont {Goncharov}(2002)}]{goncharov_prl_2002}%
  \BibitemOpen
  \bibfield  {author} {\bibinfo {author} {\bibfnamefont {V.~N.}\ \bibnamefont
  {Goncharov}},\ }\bibfield  {title} {\enquote {\bibinfo {title} {{Analytical
  Model of Nonlinear, Single-Mode, Classical Rayleigh-Taylor Instability at
  Arbitrary Atwood Numbers}},}\ }\href@noop {} {\bibfield  {journal} {\bibinfo
  {journal} {Phys. Rev. Lett.}\ }\textbf {\bibinfo {volume} {88}},\ \bibinfo
  {pages} {134502} (\bibinfo {year} {2002})}\BibitemShut {NoStop}%
\bibitem [{\citenamefont {Young}\ and\ \citenamefont
  {Ham}(2006)}]{young_jt_2006}%
  \BibitemOpen
  \bibfield  {author} {\bibinfo {author} {\bibfnamefont {Y.-N.}\ \bibnamefont
  {Young}}\ and\ \bibinfo {author} {\bibfnamefont {F.~E.}\ \bibnamefont
  {Ham}},\ }\bibfield  {title} {\enquote {\bibinfo {title} {{Surface tension in
  incompressible Rayleigh-Taylor mixing flow}},}\ }\href@noop {} {\bibfield
  {journal} {\bibinfo  {journal} {J. Turbulence}\ }\textbf {\bibinfo {volume}
  {7}},\ \bibinfo {pages} {1--23} (\bibinfo {year} {2006})}\BibitemShut
  {NoStop}%
\bibitem [{\citenamefont {Sohn}(2009)}]{sohn_prE_2009}%
  \BibitemOpen
  \bibfield  {author} {\bibinfo {author} {\bibfnamefont {S.-I.}\ \bibnamefont
  {Sohn}},\ }\bibfield  {title} {\enquote {\bibinfo {title} {{Effects of
  surface tension and viscosity on the growth rates of Rayleigh-Taylor and
  Richtmyer-Meshkov instabilities}},}\ }\href@noop {} {\bibfield  {journal}
  {\bibinfo  {journal} {Phys. Rev. E}\ }\textbf {\bibinfo {volume} {80}},\
  \bibinfo {pages} {055302} (\bibinfo {year} {2009})}\BibitemShut {NoStop}%
\bibitem [{\citenamefont {Ramaprabhu}\ \emph {et~al.}(2006)\citenamefont
  {Ramaprabhu}, \citenamefont {Dimonte}, \citenamefont {Young}, \citenamefont
  {Calder},\ and\ \citenamefont {Fryxell}}]{ramaprabhu_prE_2006}%
  \BibitemOpen
  \bibfield  {author} {\bibinfo {author} {\bibfnamefont {P.}~\bibnamefont
  {Ramaprabhu}}, \bibinfo {author} {\bibfnamefont {G.}~\bibnamefont {Dimonte}},
  \bibinfo {author} {\bibfnamefont {Y.-N.}\ \bibnamefont {Young}}, \bibinfo
  {author} {\bibfnamefont {A.~C.}\ \bibnamefont {Calder}}, \ and\ \bibinfo
  {author} {\bibfnamefont {B.}~\bibnamefont {Fryxell}},\ }\bibfield  {title}
  {\enquote {\bibinfo {title} {{Limits of the potential flow approach to the
  single-mode Rayleigh-Taylor problem}},}\ }\href@noop {} {\bibfield  {journal}
  {\bibinfo  {journal} {Phys. Rev. E}\ }\textbf {\bibinfo {volume} {74}},\
  \bibinfo {pages} {066308} (\bibinfo {year} {2006})}\BibitemShut {NoStop}%
\bibitem [{\citenamefont {Wei}\ and\ \citenamefont
  {Livescu}(2012)}]{wei_prE_2012}%
  \BibitemOpen
  \bibfield  {author} {\bibinfo {author} {\bibfnamefont {T.}~\bibnamefont
  {Wei}}\ and\ \bibinfo {author} {\bibfnamefont {D.}~\bibnamefont {Livescu}},\
  }\bibfield  {title} {\enquote {\bibinfo {title} {{Late-time quadratic growth
  in single-mode Rayleigh-Taylor instability}},}\ }\href@noop {} {\bibfield
  {journal} {\bibinfo  {journal} {Phys. Rev. E}\ }\textbf {\bibinfo {volume}
  {86}},\ \bibinfo {pages} {046405} (\bibinfo {year} {2012})}\BibitemShut
  {NoStop}%
\bibitem [{\citenamefont {Ramaprabhu}\ \emph {et~al.}(2012)\citenamefont
  {Ramaprabhu}, \citenamefont {Dimonte}, \citenamefont {Woodward},
  \citenamefont {Fryer}, \citenamefont {Rockefeller}, \citenamefont
  {Muthuraman}, \citenamefont {Lin},\ and\ \citenamefont
  {Jayaraj}}]{ramaprabhu_pof_2012}%
  \BibitemOpen
  \bibfield  {author} {\bibinfo {author} {\bibfnamefont {P.}~\bibnamefont
  {Ramaprabhu}}, \bibinfo {author} {\bibfnamefont {G.}~\bibnamefont {Dimonte}},
  \bibinfo {author} {\bibfnamefont {P.}~\bibnamefont {Woodward}}, \bibinfo
  {author} {\bibfnamefont {C.}~\bibnamefont {Fryer}}, \bibinfo {author}
  {\bibfnamefont {G.}~\bibnamefont {Rockefeller}}, \bibinfo {author}
  {\bibfnamefont {K.}~\bibnamefont {Muthuraman}}, \bibinfo {author}
  {\bibfnamefont {P.-H.}\ \bibnamefont {Lin}}, \ and\ \bibinfo {author}
  {\bibfnamefont {J.}~\bibnamefont {Jayaraj}},\ }\bibfield  {title} {\enquote
  {\bibinfo {title} {{The late-time dynamics of the single-mode Rayleigh-Taylor
  instability}},}\ }\href@noop {} {\bibfield  {journal} {\bibinfo  {journal}
  {Phys. Fluids}\ }\textbf {\bibinfo {volume} {24}},\ \bibinfo {pages} {074107}
  (\bibinfo {year} {2012})}\BibitemShut {NoStop}%
\bibitem [{\citenamefont {Hu}\ \emph {et~al.}(2019)\citenamefont {Hu},
  \citenamefont {Zhang}, \citenamefont {Tian}, \citenamefont {He},\ and\
  \citenamefont {Li}}]{hu_pof_2019}%
  \BibitemOpen
  \bibfield  {author} {\bibinfo {author} {\bibfnamefont {Z.-X.}\ \bibnamefont
  {Hu}}, \bibinfo {author} {\bibfnamefont {Y.-S.}\ \bibnamefont {Zhang}},
  \bibinfo {author} {\bibfnamefont {B.}~\bibnamefont {Tian}}, \bibinfo {author}
  {\bibfnamefont {Z.}~\bibnamefont {He}}, \ and\ \bibinfo {author}
  {\bibfnamefont {L.}~\bibnamefont {Li}},\ }\bibfield  {title} {\enquote
  {\bibinfo {title} {{Effect of viscosity on two-dimensional single-mode
  Rayleigh-Taylor instability during and after the reacceleration stage}},}\
  }\href@noop {} {\bibfield  {journal} {\bibinfo  {journal} {Phys. Fluids}\
  }\textbf {\bibinfo {volume} {31}},\ \bibinfo {pages} {104108} (\bibinfo
  {year} {2019})}\BibitemShut {NoStop}%
\bibitem [{\citenamefont {Anderson}, \citenamefont {McFadden},\ and\
  \citenamefont {Wheeler}(1998)}]{anderson_arfm_1998}%
  \BibitemOpen
  \bibfield  {author} {\bibinfo {author} {\bibfnamefont {D.~M.}\ \bibnamefont
  {Anderson}}, \bibinfo {author} {\bibfnamefont {G.~B.}\ \bibnamefont
  {McFadden}}, \ and\ \bibinfo {author} {\bibfnamefont {A.~A.}\ \bibnamefont
  {Wheeler}},\ }\bibfield  {title} {\enquote {\bibinfo {title}
  {{Diffuse-interface methods in fluid mechanics}},}\ }\href@noop {} {\bibfield
   {journal} {\bibinfo  {journal} {Annu. Rev. Fluid Mech.}\ }\textbf {\bibinfo
  {volume} {30}},\ \bibinfo {pages} {139--165} (\bibinfo {year}
  {1998})}\BibitemShut {NoStop}%
\bibitem [{\citenamefont {Cahn}\ and\ \citenamefont
  {Hilliard}(1958)}]{cahn_jchp_1958}%
  \BibitemOpen
  \bibfield  {author} {\bibinfo {author} {\bibfnamefont {J.~W.}\ \bibnamefont
  {Cahn}}\ and\ \bibinfo {author} {\bibfnamefont {J.~E.}\ \bibnamefont
  {Hilliard}},\ }\bibfield  {title} {\enquote {\bibinfo {title} {{Free Energy
  of a Nonuniform System. I. Interfacial Free Energy}},}\ }\href@noop {}
  {\bibfield  {journal} {\bibinfo  {journal} {J. Chem. Phys.}\ }\textbf
  {\bibinfo {volume} {28}},\ \bibinfo {pages} {258--267} (\bibinfo {year}
  {1958})}\BibitemShut {NoStop}%
\bibitem [{\citenamefont {Cahn}(1965)}]{cahn_jchp_1965}%
  \BibitemOpen
  \bibfield  {author} {\bibinfo {author} {\bibfnamefont {J.~W.}\ \bibnamefont
  {Cahn}},\ }\bibfield  {title} {\enquote {\bibinfo {title} {{Phase Separation
  by Spinodal Decomposition in Isotropic Systems}},}\ }\href@noop {} {\bibfield
   {journal} {\bibinfo  {journal} {J. Chem. Phys.}\ }\textbf {\bibinfo {volume}
  {42}},\ \bibinfo {pages} {93--99} (\bibinfo {year} {1965})}\BibitemShut
  {NoStop}%
\bibitem [{\citenamefont {Magaletti}\ \emph {et~al.}(2013)\citenamefont
  {Magaletti}, \citenamefont {Picano}, \citenamefont {Chinappi}, \citenamefont
  {Marino},\ and\ \citenamefont {Casciola}}]{magaletti_jfm_2013}%
  \BibitemOpen
  \bibfield  {author} {\bibinfo {author} {\bibfnamefont {F.}~\bibnamefont
  {Magaletti}}, \bibinfo {author} {\bibfnamefont {F.}~\bibnamefont {Picano}},
  \bibinfo {author} {\bibfnamefont {M.}~\bibnamefont {Chinappi}}, \bibinfo
  {author} {\bibfnamefont {L.}~\bibnamefont {Marino}}, \ and\ \bibinfo {author}
  {\bibfnamefont {C.~M.}\ \bibnamefont {Casciola}},\ }\bibfield  {title}
  {\enquote {\bibinfo {title} {{The sharp-interface limit of the
  Cahn-Hilliard/Navier-Stokes model for binary fluids}},}\ }\href@noop {}
  {\bibfield  {journal} {\bibinfo  {journal} {J. Fluid Mech.}\ }\textbf
  {\bibinfo {volume} {714}},\ \bibinfo {pages} {95--126} (\bibinfo {year}
  {2013})}\BibitemShut {NoStop}%
\bibitem [{\citenamefont {Celani}\ \emph {et~al.}(2009)\citenamefont {Celani},
  \citenamefont {Mazzino}, \citenamefont {Muratore-Ginanneschi},\ and\
  \citenamefont {Vozella}}]{celani_jfm_2009}%
  \BibitemOpen
  \bibfield  {author} {\bibinfo {author} {\bibfnamefont {A.}~\bibnamefont
  {Celani}}, \bibinfo {author} {\bibfnamefont {A.}~\bibnamefont {Mazzino}},
  \bibinfo {author} {\bibfnamefont {P.}~\bibnamefont {Muratore-Ginanneschi}}, \
  and\ \bibinfo {author} {\bibfnamefont {L.}~\bibnamefont {Vozella}},\
  }\bibfield  {title} {\enquote {\bibinfo {title} {{Phase-field model for the
  Rayleigh-Taylor instability of immiscible fluids}},}\ }\href@noop {}
  {\bibfield  {journal} {\bibinfo  {journal} {J. Fluid Mech.}\ }\textbf
  {\bibinfo {volume} {622}},\ \bibinfo {pages} {115--134} (\bibinfo {year}
  {2009})}\BibitemShut {NoStop}%
\bibitem [{\citenamefont {Lee}, \citenamefont {Kim},\ and\ \citenamefont
  {Kim}(2011)}]{lee_ijnme_2011}%
  \BibitemOpen
  \bibfield  {author} {\bibinfo {author} {\bibfnamefont {H.~G.}\ \bibnamefont
  {Lee}}, \bibinfo {author} {\bibfnamefont {K.}~\bibnamefont {Kim}}, \ and\
  \bibinfo {author} {\bibfnamefont {J.}~\bibnamefont {Kim}},\ }\bibfield
  {title} {\enquote {\bibinfo {title} {{On the long time simulation of the
  Rayleigh–Taylor instability}},}\ }\href@noop {} {\bibfield  {journal}
  {\bibinfo  {journal} {Int. J. Numer. Meth. Engng}\ }\textbf {\bibinfo
  {volume} {85}},\ \bibinfo {pages} {1633--1647} (\bibinfo {year}
  {2011})}\BibitemShut {NoStop}%
\bibitem [{\citenamefont {Lee}\ and\ \citenamefont {Kim}(2013)}]{lee_cma_2013}%
  \BibitemOpen
  \bibfield  {author} {\bibinfo {author} {\bibfnamefont {H.~G.}\ \bibnamefont
  {Lee}}\ and\ \bibinfo {author} {\bibfnamefont {J.}~\bibnamefont {Kim}},\
  }\bibfield  {title} {\enquote {\bibinfo {title} {{Numerical simulation of the
  three-dimensional Rayleigh-Taylor instability}},}\ }\href@noop {} {\bibfield
  {journal} {\bibinfo  {journal} {Computer and Mathematics with Applications}\
  }\textbf {\bibinfo {volume} {66}},\ \bibinfo {pages} {1466--1474} (\bibinfo
  {year} {2013})}\BibitemShut {NoStop}%
\bibitem [{\citenamefont {Hong}\ \emph {et~al.}(2019)\citenamefont {Hong},
  \citenamefont {Xiaoliang}, \citenamefont {Xuefeng},\ and\ \citenamefont
  {Jiangrong}}]{hong2019}%
  \BibitemOpen
  \bibfield  {author} {\bibinfo {author} {\bibfnamefont {L.}~\bibnamefont
  {Hong}}, \bibinfo {author} {\bibfnamefont {H.}~\bibnamefont {Xiaoliang}},
  \bibinfo {author} {\bibfnamefont {H.}~\bibnamefont {Xuefeng}}, \ and\
  \bibinfo {author} {\bibfnamefont {X.}~\bibnamefont {Jiangrong}},\ }\bibfield
  {title} {\enquote {\bibinfo {title} {{Direct numerical simulations of
  multi-mode immiscible Rayleigh-Taylor instability with high Reynolds
  numbers}},}\ }\href@noop {} {\bibfield  {journal} {\bibinfo  {journal} {Phys.
  Fluids}\ }\textbf {\bibinfo {volume} {31}},\ \bibinfo {pages} {112104}
  (\bibinfo {year} {2019})}\BibitemShut {NoStop}%
\bibitem [{\citenamefont {Nitschke}, \citenamefont {Voigt},\ and\ \citenamefont
  {Wensch}(2012)}]{nitschke_jfm_2012}%
  \BibitemOpen
  \bibfield  {author} {\bibinfo {author} {\bibfnamefont {I.}~\bibnamefont
  {Nitschke}}, \bibinfo {author} {\bibfnamefont {A.}~\bibnamefont {Voigt}}, \
  and\ \bibinfo {author} {\bibfnamefont {J.}~\bibnamefont {Wensch}},\
  }\bibfield  {title} {\enquote {\bibinfo {title} {{A finite element approach
  to incompressible two-phase flow on manifolds}},}\ }\href@noop {} {\bibfield
  {journal} {\bibinfo  {journal} {J. Fluid Mech.}\ }\textbf {\bibinfo {volume}
  {708}},\ \bibinfo {pages} {418--438} (\bibinfo {year} {2012})}\BibitemShut
  {NoStop}%
\bibitem [{\citenamefont {Yang}\ \emph {et~al.}(2016)\citenamefont {Yang},
  \citenamefont {Li}, \citenamefont {Zhao}, \citenamefont {Shao},\ and\
  \citenamefont {Xu}}]{yang2016}%
  \BibitemOpen
  \bibfield  {author} {\bibinfo {author} {\bibfnamefont {Q.}~\bibnamefont
  {Yang}}, \bibinfo {author} {\bibfnamefont {B.~Q.}\ \bibnamefont {Li}},
  \bibinfo {author} {\bibfnamefont {Z.}~\bibnamefont {Zhao}}, \bibinfo {author}
  {\bibfnamefont {J.}~\bibnamefont {Shao}}, \ and\ \bibinfo {author}
  {\bibfnamefont {F.}~\bibnamefont {Xu}},\ }\bibfield  {title} {\enquote
  {\bibinfo {title} {{Numerical analysis of the Rayleigh-Taylor instability in
  an electric field}},}\ }\href@noop {} {\bibfield  {journal} {\bibinfo
  {journal} {J. Fluid Mech.}\ }\textbf {\bibinfo {volume} {792}},\ \bibinfo
  {pages} {397--434} (\bibinfo {year} {2016})}\BibitemShut {NoStop}%
\bibitem [{\citenamefont {Lyubimova}, \citenamefont {Vorobev},\ and\
  \citenamefont {Prokopev}(2019)}]{lyubimova2019}%
  \BibitemOpen
  \bibfield  {author} {\bibinfo {author} {\bibfnamefont {T.}~\bibnamefont
  {Lyubimova}}, \bibinfo {author} {\bibfnamefont {A.}~\bibnamefont {Vorobev}},
  \ and\ \bibinfo {author} {\bibfnamefont {S.}~\bibnamefont {Prokopev}},\
  }\bibfield  {title} {\enquote {\bibinfo {title} {{Rayleigh-Taylor instability
  of a miscible interface in a confined domain}},}\ }\href@noop {} {\bibfield
  {journal} {\bibinfo  {journal} {Phys. Fluids}\ }\textbf {\bibinfo {volume}
  {31}},\ \bibinfo {pages} {014104} (\bibinfo {year} {2019})}\BibitemShut
  {NoStop}%
\bibitem [{\citenamefont {Song}\ \emph {et~al.}(2019)\citenamefont {Song},
  \citenamefont {Plana}, \citenamefont {Lopez},\ and\ \citenamefont
  {Avila}}]{song2019}%
  \BibitemOpen
  \bibfield  {author} {\bibinfo {author} {\bibfnamefont {B.}~\bibnamefont
  {Song}}, \bibinfo {author} {\bibfnamefont {C.}~\bibnamefont {Plana}},
  \bibinfo {author} {\bibfnamefont {J.~M.}\ \bibnamefont {Lopez}}, \ and\
  \bibinfo {author} {\bibfnamefont {M.}~\bibnamefont {Avila}},\ }\bibfield
  {title} {\enquote {\bibinfo {title} {{Phase-field simulation of core-annular
  pipe flow}},}\ }\href@noop {} {\bibfield  {journal} {\bibinfo  {journal}
  {Int. J. Multiphase Flow}\ }\textbf {\bibinfo {volume} {117}},\ \bibinfo
  {pages} {14--24} (\bibinfo {year} {2019})}\BibitemShut {NoStop}%
\bibitem [{\citenamefont {Kendon}\ \emph {et~al.}(2001)\citenamefont {Kendon},
  \citenamefont {Cates}, \citenamefont {Desplat}, \citenamefont
  {Pagonabarraga},\ and\ \citenamefont {Bladon}}]{kendon_jfm_2001}%
  \BibitemOpen
  \bibfield  {author} {\bibinfo {author} {\bibfnamefont {V.~M.}\ \bibnamefont
  {Kendon}}, \bibinfo {author} {\bibfnamefont {M.~E.}\ \bibnamefont {Cates}},
  \bibinfo {author} {\bibfnamefont {J.-C.}\ \bibnamefont {Desplat}}, \bibinfo
  {author} {\bibfnamefont {I.}~\bibnamefont {Pagonabarraga}}, \ and\ \bibinfo
  {author} {\bibfnamefont {P.}~\bibnamefont {Bladon}},\ }\bibfield  {title}
  {\enquote {\bibinfo {title} {{Inertial effects in three dimensional spinodal
  decomposition of a symmetric binary fluid mixture: A lattice Boltzmann
  study}},}\ }\href@noop {} {\bibfield  {journal} {\bibinfo  {journal} {J.
  Fluid Mech.}\ }\textbf {\bibinfo {volume} {440}},\ \bibinfo {pages}
  {147--203} (\bibinfo {year} {2001})}\BibitemShut {NoStop}%
\bibitem [{\citenamefont {Henry}\ and\ \citenamefont
  {Tegze}(2018)}]{Henry2018}%
  \BibitemOpen
  \bibfield  {author} {\bibinfo {author} {\bibfnamefont {H.}~\bibnamefont
  {Henry}}\ and\ \bibinfo {author} {\bibfnamefont {G.}~\bibnamefont {Tegze}},\
  }\bibfield  {title} {\enquote {\bibinfo {title} {{Self-similarity and
  coarsening rate of a convecting bicontinuous phase separating mixture :
  Effect of the viscosity contrast}},}\ }\href {\doibase
  10.1103/PhysRevFluids.3.074306} {\bibfield  {journal} {\bibinfo  {journal}
  {Phys. Rev. F}\ }\textbf {\bibinfo {volume} {3}},\ \bibinfo {pages} {074306}
  (\bibinfo {year} {2018})}\BibitemShut {NoStop}%
\bibitem [{\citenamefont {Henry}\ and\ \citenamefont
  {Tegze}(2019)}]{Henry2019}%
  \BibitemOpen
  \bibfield  {author} {\bibinfo {author} {\bibfnamefont {H.}~\bibnamefont
  {Henry}}\ and\ \bibinfo {author} {\bibfnamefont {G.}~\bibnamefont {Tegze}},\
  }\bibfield  {title} {\enquote {\bibinfo {title} {{Kinetics of coarsening have
  dramatic effects on the microstructure : Self-similarity breakdown induced by
  viscosity contrast}},}\ }\href {\doibase 10.1103/PhysRevE.100.013116}
  {\bibfield  {journal} {\bibinfo  {journal} {Phys. Rev. E}\ }\textbf {\bibinfo
  {volume} {100}},\ \bibinfo {pages} {013116} (\bibinfo {year}
  {2019})}\BibitemShut {NoStop}%
\end{thebibliography}%

\end{document}